\documentclass[a4paper, 11pt]{article}

\usepackage[papersize={210mm,297mm},top=30mm, bottom=30mm, left=26mm , right=26mm]{geometry}

\usepackage{listings}
\usepackage[usenames,dvipsnames]{xcolor}

\usepackage[colorlinks,citecolor=magenta,urlcolor=blue]{hyperref}

\RequirePackage{amsmath, amssymb}

\DeclareMathOperator*{\argmin}{argmin}

\usepackage{algorithm}
\usepackage{algorithmic}
\usepackage{soul}
\usepackage{tikz}
\usetikzlibrary{arrows}
\usepackage{expl3}
\ExplSyntaxOn
\cs_set_eq:NN \fpeval \fp_eval:n
\ExplSyntaxOff

\usepackage{texments}
\usepackage{natbib}

\let\proglang=\textsf

\RequirePackage{graphicx,color,ae,fancyvrb}
\DefineVerbatimEnvironment{Sinput}{Verbatim}{fontshape=sl}
\DefineVerbatimEnvironment{Soutput}{Verbatim}{}
\DefineVerbatimEnvironment{Scode}{Verbatim}{fontshape=sl}

\DefineVerbatimEnvironment{Code}{Verbatim}{}
\DefineVerbatimEnvironment{CodeInput}{Verbatim}{fontshape=sl}
\DefineVerbatimEnvironment{CodeOutput}{Verbatim}{}
\newenvironment{CodeChunk}{}{}
\setkeys{Gin}{width=0.8\textwidth}

\newcommand{\fct}[1]{\code{#1()}}

\newcommand{\pkg}[1]{{\fontseries{b}\selectfont #1}}

\begin{document}

\title{\proglang{gfpop}: an \proglang{R} Package for Univariate Graph-Constrained Change-point Detection}

\author{
 Vincent Runge\\Universit\'e d'\'Evry\footnote{E-mail: vincent.runge@univ-evry.fr}
  \and
Toby Dylan Hocking\\ Northern Arizona University
\and
Gaetano Romano\\Lancaster University
\and 
Fatemeh Afghah\\Northern Arizona University
\and 
Paul Fearnhead\\Lancaster University
\and
Guillem Rigaill\\INRAE - Universit\'e d'\'Evry
}


\date{}
\maketitle

\begin{abstract}
In a world with data that change rapidly and abruptly, it is important to detect those changes accurately. In this paper we describe an \proglang{R} package implementing a generalized version of an algorithm recently proposed by \cite{hocking2020constrained} for penalized maximum likelihood inference of constrained multiple change-point models. This algorithm can be used to pinpoint the precise locations of abrupt changes in large data sequences. 
There are many application domains for such models, such as medicine, neuroscience or genomics. Often, practitioners have prior knowledge about the changes they are looking for. For example in genomic data, biologists sometimes expect peaks: up changes followed by down changes. Taking advantage of such prior information can substantially improve the accuracy with which we can detect and estimate changes. \cite{hocking2020constrained} described a graph framework to encode many examples of such prior information and a generic algorithm to infer the optimal model parameters, but implemented the algorithm for just a single scenario. 
We present the \pkg{gfpop} package that implements the algorithm in a generic manner in \proglang{R/C++}. \pkg{gfpop} works for a user-defined graph that can encode prior assumptions about the types of change that are possible and implements several loss functions (Gauss, Poisson, binomial, biweight and Huber). 
We then illustrate the use of \pkg{gfpop} on isotonic simulations and several applications in biology. For a number of graphs the algorithm runs in a matter of seconds or minutes for $10^5$ data points.
\end{abstract}



\section{Introduction}

\subsection[Multiple change-point R packages]{Multiple change-point \proglang{R} packages}

In the last decade there has been an increasing interest in algorithms for detecting changes in mean. There are a variety of approaches to detecting such change-points, see \citet{truong2020selective}  for a recent review of the area. Many of these recursively apply a test for a single change-point. These include binary segmentation \citep{scott1974cluster} and its variants \citep{olshen2004circular,fryzlewicz2014wild},  multiscale methods \citep{frick2014multiscale} and MOSUM methods \citep{eichinger2018mosum}. \proglang{R} packages that implement these and related methods include \pkg{wbs} \citep{baranowski2014wbs}, \pkg{not} \citep{baranowski2019narrowest}, \pkg{breakfast} \citep{breakfast}, \pkg{stepR} \citep{stepR}, and \pkg{mosum} \citep{meier2019mosum}. See \cite{fearnhead2020relating} for a comparison of many of these methods.

Alternatively one can try to jointly estimate the location of all change-points by maximizing a penalized likelihood or, equivalently, minimizing a penalized cost. Dynamic programming was originally proposed in the change-point literature in the context of the ``segment neighborhood'' (SN) and ``optimal partitioning'' (OP) algorithms \citep{auger1989algorithms,jackson2005algorithm}. 
More recently \citet{killick2012optimal} proposed the PELT pruning rules, which reduces time complexity from quadratic to linear in asymptotic regimes where the number of change-points increases as we observe more data. This work has stimulated a new interest in these problems.  The \proglang{R} package \pkg{changepoint} \citep{killick2014changepoint} and \pkg{changepoint.np} \citep{haynes2016changepoint} are based on these PELT \emph{inequality} pruning rules. 
A new \emph{functional} pruning rule was independently discovered by \citet{johnson2013} and \citet{rigaill2015pruned}. 
When comparing these two pruning rules, the \emph{functional} pruning always prunes more than PELT \emph{inequality} pruning \cite[see Theorem~2 and Figures~4 and~5 in][]{maidstone2017optimal}. Furthermore {functional} pruning  empirically shows reduced time complexity in many situations. For example, when we have data with no changes, PELT pruning algorithms have a quadratic complexity, but functional pruning algorithms can have a log-linear complexity \cite[see Section 7 in][]{maidstone2017optimal}. Functional pruning algorithms are implemented in \proglang{R} packages \pkg{fpop}, \pkg{Segmentor3IsBack} \citep{cleynen2014segmentor3isback} and \pkg{jointseg} \citep{jointseg}.

Besides time efficiency, recent efforts have been made to extend the class of change-point models considered by adding constraints. For example, in applications which involve detecting peaks in genomic data, the inference is constrained to return a sequence of down and up segments by packages \pkg{PeakSegDP} \citep{hocking2015peakseg}, \pkg{PeakSegOptimal} \citep{hocking2020constrained}, \pkg{PeakSegDisk} \citep{JSSv101i10}, and \pkg{PeakSegJoint} \citep{Hocking2020psb}. Including such constraints, when appropriate for the application, has been shown to substantially improve the accuracy of change-point detection \citep{hocking2020constrained}.
Whereas these previous packages only implement up/down constraints and the Poisson loss, the proposed \pkg{gfpop} package is the first to implement inference for a wide range of constraints, allows specifying models that can mix different types of changes through a graph, and also allows for other loss functions. For that reason we named our package \pkg{gfpop} as an abbreviation for ``generalized functional pruning optimal partitioning''.
Our intention is to provide a user-friendly package which popularizes these recent discoveries about the functional pruning method, by allowing the user to specify a wide variety of constraints and loss functions, using prior information about their data and application domain.


\subsection{Standard multiple change-point model} \label{subsec:std}

Multiple change-point models are designed to find abrupt changes in a signal. In the standard Gaussian noise model, we have data $Y_{1:n}=(Y_1,\ldots,Y_n)$ where each data point, $Y_t$, is an independent random variable with $Y_t \sim \mathcal{N}(\mu_t, \sigma ^ 2) \, $ 
and $ t \mapsto \mu_t $ is piecewise constant. The goal is to estimate the number and position of the changes, that is to find all $t$ such that $ \mu_t \neq \mu_{t + 1}$  from the observed data $(y_t)_{t = 1, ..., n}$. A classical way to proceed is to optimize the log-likelihood by fixing the number of changes. It is also possible to penalize each change by a positive penalty $\beta$  and minimize in $ \mu = (\mu_1, ..., \mu_n)^\top \in \ \mathbb{R}^n$ the following least squares criterion:
$$ Q^{std}_n(\mu) = \sum_{t = 1}^n (y_t - \mu_t)^2 + \beta \sum_ {t = 1}^{n-1} I_{\mu_t \neq \mu_{t + 1}} \ , $$
where $I \in \{0,1\}$ is the indicator function ($I_x=1$ if $x$ is true, and is zero otherwise). In both cases, fast dynamic programming algorithms can solve the related optimization problem exactly \citep{killick2012optimal, rigaill2015pruned, maidstone2017optimal}. In Section \ref{sec:opandupdate} we derive the explicit form of the simpler FPOP update-rule for our more general graph framework.

\subsection{Constrained multiple change-point model}

In many applications, it is desirable to constrain the parameters of successive segments \citep{hocking2015peakseg, maidstone2017optimal,jewell2018fast,baranowski2014wbs}. This means that the $\mu$ parameter is restricted by inequalities to a subset of $\mathbb{R}^n$. Arguably, the simplest and most studied case is isotonic regression \citep{barlow1972}. In this case the goal is to minimize in $\mu$ the constrained least-squares criterion:
$$Q^{iso}_n(\mu) = \sum_{t = 1}^n (y_t - \mu_t)^2 \ , \ 
          \hbox{subject to the constraint} \ \mu_t \leq \mu_{t + 1},\,\, \forall \ t \in \{1,...,n-1\}. $$
The obtained estimator is piecewise constant, which makes the link with the multiple change-point problem. 
Several efficient inference algorithms have been proposed \citep{best1990active, johnson2013, gao2017estimation}, and the \pkg{isotone} package provide an implementation \citep{de2010isotone}.

More generally, we may want to impose more complex patterns, such as unimodality \citep{stout2008unimodal} or a succession of up and down changes \citep{hocking2015peakseg} to detect peaks. There are very efficient algorithms for the isotonic and unimodal cases \citep{best1990active,stout2008unimodal} at least if the number of changes is not penalized. 
For more complex constraints like the up-down pattern, \cite{hocking2020constrained} proposed an exact algorithm. 
This algorithm is a generalization of the functional dynamic programming algorithm of \cite{rigaill2015pruned} and \cite{maidstone2017optimal}. 
Variants of this algorithm allow penalizing or constraining the number of changes. 
Other variants allow robust losses, including the biweight loss, instead of the least-squares criterion for assessing fit to the data. 
In the case of non-constrained (standard) multiple change-point detection, the biweight loss has good statistical properties \citep{fearnhead2018changepoint}. 
The simulations of \cite{bach2018efficient} in the context of isotonic regression also show the benefit of such losses. 


\subsection{Contributions}

\cite{hocking2020constrained} described a graph-based framework to encode prior constraints on how parameters change at each change-point, and a generic algorithm to infer the optimal model parameters. 
However, they implemented the algorithm for a single scenario (Poisson loss and up/down constraints). 
The \pkg{gfpop} package implements their algorithm in a generic manner in \proglang{R/C++}, for user-defined graphs and several loss functions.

\subsection{Outline}

In Section \ref{sec:constr} we formally define the graphs and explain their connection to HMM. We also provide numerous graph examples to illustrate the versatility of our framework. In Section \ref{sec:optim} we present the optimization problem solved by our package. In Section \ref{sec:howto} we go through the main functions of the package. We illustrate in Section \ref{sec:examples} the use of our package on various real data sets. Finally, using simulations we compare in Section \ref{sec:iso} the result of our package with those of standard isotonic packages and show the benefit of using robust losses and penalizing the number of segments.


\section{Constraint graphs and change-points model as a HMM}\label{sec:constr}

We begin by recasting the standard and constrained multiple change-point problem as a continuous hidden Markov model (HMM) with a particular transition kernel represented as a graph \citep{johnson2013}.
At each time $t$ the signal can be in a number of states, which are nodes of the graph.
Possible transitions between states at time $t$ and $t + 1$ are represented by the edges of the graph. 
Each edge has three properties: a constraint (e.g., $ \mu_t \leq \mu_{t + 1}$), a penalty (possibly null) and a loss function (cost associated to a data-point).
In \pkg{gfpop} the set of transitions is constant over time, leading to a collapsed representation of the graph. 
We then formalize the concept of a valid signal or path, that is one satisfying all constraints, and finally present a number of examples.

\subsection{Transition kernel and graph of constraints}\label{sec:HMM}

\begin{figure}[!htbp]
\centering{
\begin{tikzpicture}[>=latex']
 \node[draw,circle] at (0, 0) (Z_1) {$Z_1$}; \node[draw,circle] at (2, 0) (Z_2) {$Z_2$};
 \node[draw,circle] at (4, 0) (Z_3) {$Z_3$}; \node[draw,circle] at (6, 0) (Z_4) {$Z_4$};
 \node[draw,circle] at (8, 0) (Z_5) {$Z_n$};
 
 \node[draw,circle] at (0, -1.5) (Y_1) {$Y_1$};
 \node[draw,circle] at (2, -1.5) (Y_2) {$Y_2$};
 \node[draw,circle] at (4, -1.5) (Y_3) {$Y_3$};
 \node[draw,circle] at (6, -1.5) (Y_4) {$Y_4$};
 \node[draw,circle] at (8, -1.5) (Y_5) {$Y_n$};

 \draw [->] (Z_1) to (Z_2); \draw [->] (Z_2) to (Z_3); \draw [->] (Z_3) to (Z_4);
 \draw [->, dotted] (Z_4) to (Z_5);
 \draw [->] (Z_1) to (Y_1); \draw [->] (Z_2) to (Y_2); \draw [->] (Z_3) to (Y_3);
 \draw [->] (Z_4) to (Y_4);
  \draw [->] (Z_5) to (Y_5);
\end{tikzpicture}
}
\caption{Multiple change-point model as a Hidden Markov Model.}
\label{fig:HMM}
\end{figure}
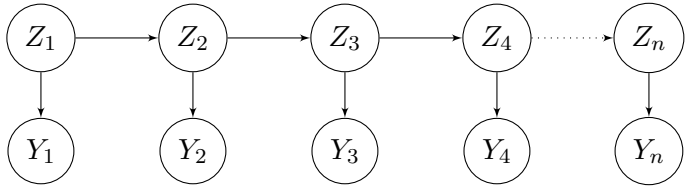

\paragraph{Standard multiple change-point model as a HMM.}
It is possible to recast the classic multiple change-point model as a Hidden Markov Model with a continuous state space. Precisely, we define random variables $Z_1, ..., Z_n$ in $\mathbb{R}$ or some interval $[a, b]$. We consider a transition kernel $k(x, y) \propto \ {I}_{x=y}  \ + \ e^{-\beta} {I}_{x\neq y}$. Finally, in the Gaussian case, observations are obtained as $(Y_i|Z_i=\mu) \sim \mathcal{N}(\mu, \sigma^2)$. The Bayesian Network of this model is given in Figure~\ref{fig:HMM}.

Here the state space is $\mathbb{R}$, the set of values that the mean can take. The \code{gfpop} algorithm of \cite{hocking2020constrained} allows one to consider a more complex state space in $\mathcal{S} \times \mathbb{R}$ where $\mathcal{S}$ is a finite set. In that case the transition kernel is more complex and can be described using a graph. Below, we first present the graph and then explain how it is linked to the transition kernel.


\paragraph{Graph of constraints.}\label{subsec:graphofconstraints}
The graph of constraints $\mathcal{G}_n $ is an acyclic directed graph defined as follows.

\begin{enumerate}
\item Nodes are indexed by time $t$ $ \in \{1, ..., n \} $ and a state $ s \in \mathcal{S} = \{1, ... S \} $;

\item We include two undefined states $\#, \emptyset$ for the starting nodes, $ v_0 = (0, \#) $, and arrival nodes, $v_{n + 1} = (n + 1, \emptyset)$; 

\item Edges are transitions between consecutive ``time'' nodes of type $v = (t, s)$ and $v '=(t + 1, s')$. Edges $e$ are then described by a triplet $e=(t, s, s')$ for $t \in \{0, ..., n \} $;

\item Each edge $e=(t, s, s')$  is associated with 
  \begin{itemize}

  \item An indicator function $\mathcal I_e: \mathbb{R} \times \mathbb{R} \to \{0,1 \}$ constraining successive means\footnote{We call this parameter a mean for convenience but some models consider changes in variance or in other natural parameters.} $ \mu_t$ and $\mu_{t + 1}$. For example an edge $e$ with the corresponding indicator function $\mathcal I_e (\mu_t, \mu_{t + 1}) = I_{\mu_t \leq \mu_{t + 1}}$ ensures that means are non-decreasing; while an edge with indicator function $\mathcal I_e (\mu_t, \mu_{t + 1}) = I_{\mu_t = \mu_{t + 1}}$ would correspond to no change.
  
  \item A penalty $\beta_e \ge 0$ which is used to regularize the model (larger penalty values result in more costly change-points and thus fewer change-points in the optimal model).

  \item A loss function $\gamma_e$ for data-point $y_{t+1}$\footnote{The loss function can be edge-specific: see Figure~\ref{fig:collective} and the graph construction in Section~\ref{graphconstruction}.}.
  
  \end{itemize} 

\end{enumerate}

\paragraph{Transition kernel.}

Coming back to our HMM representation of change-point models, the transition from state $(s, \mu_t)$ to $(s', \mu_{t+1})$ (up to proportionality) is

\begin{itemize}
\item $k((s, \mu_t), (s', \mu_{t+1}))= \exp(-\beta_e) \mathcal I_e(\mu_t, \mu_{t+1}),$ if there is an edge $e=(t, s, s')$ in the graph;
\item $k((s, \mu_t), (s', \mu_{t+1}))=0$ if there is no edge $e=(t, s, s')$ in the graph.
\end{itemize}

\paragraph{Some simple examples.}
In Figures~\ref{fig:std_time} and \ref{fig:iso_time} we provide the corresponding graphs for the standard and isotonic models. Notice that the only difference is that the transitions between nodes $(t,1)$ and $(t+1,1)$ are restricted to non-decreasing means in the isotonic case.

\begin{figure}[!htbp]
\centering{\scalebox{.8}{\begin{tikzpicture}[>=latex']
 \node[draw,circle, minimum size=1.6cm] at (0, 0) (step_0) {$(0, \#)$};
 \node[draw,circle, minimum size=1.6cm] at (12, 0) (step_{n+1}) {$(n+1, \emptyset)$};
 \node[draw,circle,fill=blue!20, minimum size=1.5cm] at (3.5, 0) (step_t) {$(t, 1)$};
 \node[draw,circle,fill=blue!20, minimum size=1.5cm] at (8.5, 0) (step_{t+1}) {$(t+1, 1)$};
 \draw [->, dotted, line width=1pt] (step_0) to (step_t) ;
 \draw [->, dashed, line width=1pt] (step_t) to 
			node[midway, above=0]{$I_{\mu_t = \mu_{t+1}}$} 
		    (step_{t+1}) ;
 \draw [->, line width=1pt] (step_t) to [bend left=45] 
			node[midway, above=0]{$I_{\mu_t \neq \mu_{t+1}}$,  $\beta$} 
		    (step_{t+1}) ;
 \draw [->, dotted, line width=1pt] (step_{t+1}) to (step_{n+1}) ;

\end{tikzpicture}}}
\caption{Graph of constraints for the standard multiple change-point model. We have $\mathcal{S}=\{1\}$, the loss function is always the $\ell_2$ (Gaussian loss). The penalty is omitted when equal to zero.}
\label{fig:std_time}
\end{figure}
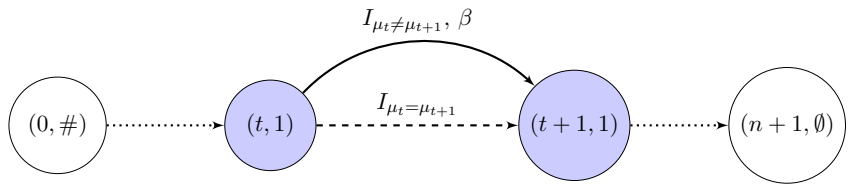

\begin{figure}[!htbp]
\centering{\scalebox{.8}{\begin{tikzpicture}[>=latex']
 \node[draw,circle, minimum size=1.6cm] at (0, 0) (step_0) {$(0, \#)$};
 \node[draw,circle, minimum size=1.6cm] at (12, 0) (step_{n+1}) {$(n+1, \emptyset)$};
 \node[draw,circle,fill=blue!20, minimum size=1.5cm] at (3.5, 0) (step_t) {$(t, 1)$};
 \node[draw,circle,fill=blue!20, minimum size=1.5cm] at (8.5, 0) (step_{t+1}) {$(t+1, 1)$};
 \draw [->, dotted,  line width=1pt] (step_0) to (step_t) ;
 \draw [->, dashed,  line width=1pt] (step_t) to 
			node[midway, above=0]{$I_{\mu_t = \mu_{t+1}}$} 
		    (step_{t+1}) ;
 \draw [->,  line width=1pt] (step_t) to [bend left=45] 
			node[midway, above=0]{$I_{\mu_t \leq \mu_{t+1}}$, $\beta$} 
		    (step_{t+1}) ;
 \draw [->, dotted,  line width=1pt] (step_{t+1}) to (step_{n+1}) ;

\end{tikzpicture}}}
\caption{Graph of constraints for the isotonic change-point model. We have $\mathcal{S}=\{1\}$, the loss function is always the $\ell_2$. The penalty is omitted when equal to zero.}
\label{fig:iso_time}
\end{figure}
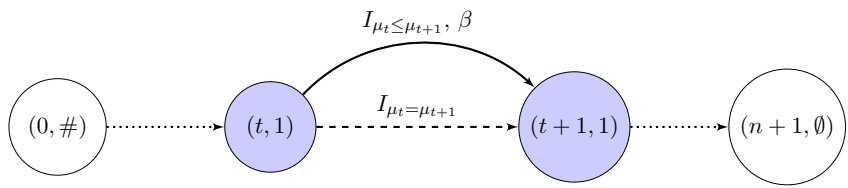

\subsection{Collapsed graph of constraints}\label{subsec:collapsed}
\begin{figure}[!htbp]
\centering{
 \begin{tikzpicture}[>=latex']
 \node[draw,circle, minimum size=1cm] at (0, 0) (step_0) {$\#$};
 \node[draw,circle, minimum size=1cm] at (8, 0) (step_{n+1}) {$\emptyset$};
 \node[draw,circle,fill=blue!20, minimum size=1cm] at (4, 0) (state_1) {$1$};
 \draw [->, dotted, line width=1pt] (step_0) to (state_1) ;
 \draw [->, dashed, loop above, line width=1pt] (state_1) to node[midway, above=0]{$I_{\mu_t = \mu_{t+1}}$} 
		    (state_1) ;
 \draw [->, loop below=0cm, line width=1pt] (state_1) to 
			node[midway, below=0]{$I_{\mu_t \neq \mu_{t+1}}$, $\beta$} 
		    (state_1) ;
 \draw [->, dotted, line width=1pt] (state_1) to (step_{n+1}) ;

\end{tikzpicture}
}
\caption{Collapsed graph of constraints for the standard multiple change-point model. 
 We have $\mathcal{S}=\{1\}$. The loss function is always the $\ell_2$. The penalty is omitted when equal to zero.}
\label{fig:std_col}
\end{figure}
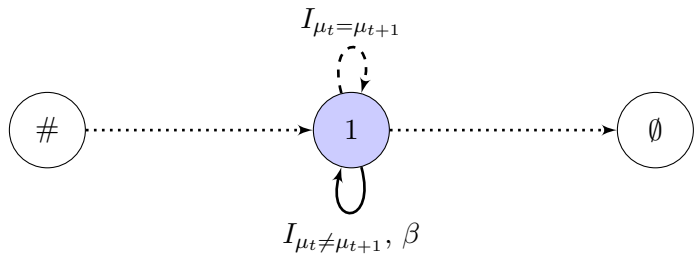

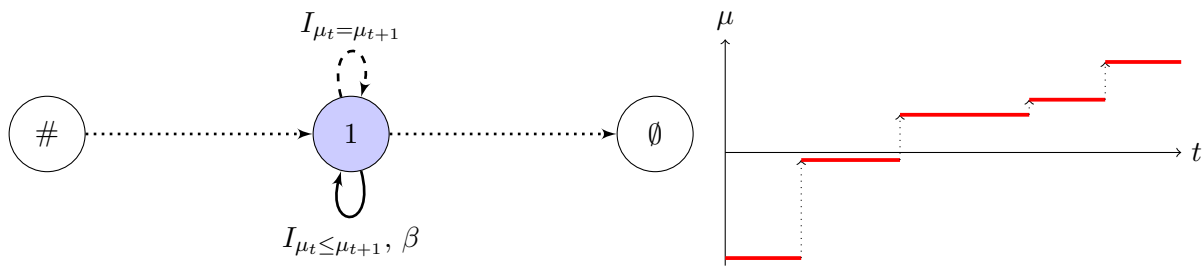
\begin{figure}[!htbp]
\begin{tikzpicture}[>=latex']
 \node[draw,circle, minimum size=1cm] at (0, 0) (step_0) {$\#$};
 \node[draw,circle, minimum size=1cm] at (8, 0) (step_{n+1}) {$\emptyset$};
 \node[draw,circle,fill=blue!20, minimum size=1cm] at (4, 0) (state_1) {$1$};
 \draw [->, dotted, line width = 1pt] (step_0) to (state_1) ;
 \draw [->, dashed, loop above, line width = 1pt] (state_1) to 
			node[midway, above=0]{$I_{\mu_t = \mu_{t+1}}$} 
		    (state_1) ;
 \draw [->, loop below, line width = 1pt] (state_1) to 
			node[midway, below=0, line width = 1pt]{$I_{\mu_t \leq \mu_{t+1}}$, $\beta$} 
		    (state_1) ;
 \draw [->, dotted, line width = 1pt] (state_1) to (step_{n+1}) ;

\end{tikzpicture}
\begin{tikzpicture}
      \draw[->] (0,0) -- (6,0) node[right] {$t$};
      \draw[->] (0,-1.5) -- (0,1.5) node[above] {$\mu$};

     \newcommand\Xa{-1.4};\newcommand\Ta{1}
     \newcommand\Xb{-0.1};\newcommand\Tb{2.3}
     \newcommand\Xc{0.5};\newcommand\Tc{4}
     \newcommand\Xd{0.7};\newcommand\Td{5}
     \newcommand\Xe{1.2};\newcommand\Te{6}
      \draw[domain=0:\Ta, line width=0.5mm,smooth,variable=\x,red] plot ({\x},\Xa);
      \draw[->,dotted] (\Ta,\Xa) -- (\Ta,\Xb);

      \draw[domain=\Ta:\Tb, line width=0.5mm,smooth,variable=\x,red] plot ({\x},\Xb);
      \draw[->,dotted] (\Tb,\Xb) -- (\Tb,\Xc);

      \draw[domain=\Tb:\Tc, line width=0.5mm,smooth,variable=\x,red] plot ({\x},\Xc);
      \draw[->,dotted] (\Tc,\Xc) -- (\Tc,\Xd);

      \draw[domain=\Tc:\Td, line width=0.5mm,smooth,variable=\x,red] plot ({\x},\Xd);
      \draw[->,dotted] (\Td,\Xd) -- (\Td,\Xe);

      \draw[domain=\Td:\Te, line width=0.5mm,smooth,variable=\x,red] plot ({\x},\Xe);
     
\end{tikzpicture}
\caption{(Left) Collapsed graph of constraints for the isotonic change-point model. We have $\mathcal{S}=\{1\}$, the loss function is always the $\ell_2$. The penalty is omitted when equal to zero.
(Right) In red, a piecewise constant function validating the graph of constraints.}
\label{fig:iso_col}
\end{figure}

In \pkg{gfpop} we only consider transitions that do not depend on time.
We then collapse the previous graph structure. To be specific, we have a single node for each $s$ and a transition from node $s$ to $s'$ if there is a transition from $(t, s)$ to $(t+1, s')$ in the full graph structure. In Figures~\ref{fig:std_col} and \ref{fig:iso_col} we provide the corresponding collapsed graphs for the standard and isotonic models. 

\paragraph{Path and constraints validation.}
In Section \ref{sec:optim} we show how we can estimate the changes by maximizing a penalized loss equal to the loss associated with the path of $\mu_t$ values plus the sum of the penalties for the edges used. This can be viewed as a maximum a-posteriori estimate based on the kernel associated with each edge and the likelihood associated with each observation.
To define this maximum properly we formalize the notion of a signal validating our constraints through the concept of a valid path in the collapsed graph.

A path $p$ of the collapsed graph $\mathcal{G}_n$ is a collection of $n + 2$ nodes $ (v_0, ..., v_ {n + 1})$ with $ v_0 = (0, \#)$, $v_{n + 1} = (n + 1, \emptyset)$ and $ v_t = (t, s_t)$ for $t \in \{1, ..., n\}$ and $s_t \in \{1, ... S\}$. In addition, the path is made of $n + 1$ edges named $e_0, ..., e_n$. Recall that each edge $e_t$ is associated to a penalty $\beta_{e_t}$, a loss $\gamma_{e_t}$ and a constraint $\mathcal I_{e_t}$. A vector $\mu \in \mathbb {R}^n$ validates the path $p$ if for all $t \in \{1, ..., n-1\}$, we have $\mathcal I_{e_t} (\mu_t, \mu_{t + 1}) = 1$ (true). We write $p(\mu) $ to say that the vector $\mu $ checks the path $p$.

The starting and arrival edges $e_0$ and $e_n$ are exceptions. For them there are neither indicator function nor associated penalty (see Figures~\ref{fig:std_time} and \ref{fig:iso_time}). However, there is a loss function $\gamma_{e_0}$ for the starting edge to add the first data-point.

\paragraph{Definition.} From now on when we use the word ``graph'' we mean the collapsed graph of constraints. In this graph, the triplet notation $(t,s,s')$ for edges is replaced by $(s,s')$. We remove the time dependency also for edges associated with starting and arrival nodes to simplify notations (even if in that case, there is a time dependency).

\subsection{A few examples}

We present a few constraint models and their graphs. Some models have been already proposed in the literature, but not necessarily using our HMM formalism. 

\begin{figure}[!htbp]
\centering{
\begin{tikzpicture}[>=latex']
 \node[draw,circle, minimum size=1cm] at (0, 0) (step_0) {$\#$};
 \node[draw,circle, minimum size=1cm] at (0, -4) (step_{n+1}) {$\emptyset$};
 \node[draw,circle,fill=blue!20, minimum size=1cm] at (11, -2) (state_Up) {Up};
 \node[draw,circle,fill=blue!20, minimum size=1cm] at (4, -2) (state_Dw) {Dw};
 
 \draw [->, dotted, line width = 1pt] (step_0) to (state_Dw) ;
 \draw [->, dashed, loop above, line width = 1pt] (state_Up) to 
			node[midway, above=0]{$I_{\mu_t = \mu_{t+1}}$} 
		    (state_Up) ;
 \draw [->, bend left, line width = 1pt] (state_Up) to 
			node[midway, below=0] {$I_{\mu_t \geq \mu_{t+1}}$, $\beta$} 
		    (state_Dw) ;

 \draw [->, dashed, loop above, line width = 1pt] (state_Dw) to 
			node[midway, above=0]{$I_{\mu_t = \mu_{t+1}}$} 
		    (state_Dw) ;
\draw [->, bend left, line width = 1pt] (state_Dw) to 
			node[midway, above=0]{$I_{\mu_t \leq \mu_{t+1}}$, $\beta$} 
		    (state_Up) ;
\draw [->, dotted, line width = 1pt] (state_Dw) to (step_{n+1}) ;
 
\end{tikzpicture}

\begin{tikzpicture}
      \draw[->] (0,0) -- (6,0) node[right] {$t$};
      \draw[->] (0,-1.5) -- (0,1.5) node[above] {$\mu$};

     \newcommand\Xa{-1.4};\newcommand\Ta{1}
     \newcommand\Xb{1.3};\newcommand\Tb{2.3}
     \newcommand\Xc{0.1};\newcommand\Tc{4}
     \newcommand\Xd{1.2};\newcommand\Td{5}
     \newcommand\Xe{-1.2};\newcommand\Te{6}
      \draw[domain=0:\Ta, line width=0.5mm,smooth,variable=\x,red] plot ({\x},\Xa);
      \draw[->,dotted] (\Ta,\Xa) -- (\Ta,\Xb);

      \draw[domain=\Ta:\Tb, line width=0.5mm,smooth,variable=\x,red] plot ({\x},\Xb);
      \draw[->,dotted] (\Tb,\Xb) -- (\Tb,\Xc);

      \draw[domain=\Tb:\Tc, line width=0.5mm,smooth,variable=\x,red] plot ({\x},\Xc);
      \draw[->,dotted] (\Tc,\Xc) -- (\Tc,\Xd);

      \draw[domain=\Tc:\Td, line width=0.5mm,smooth,variable=\x,red] plot ({\x},\Xd);
      \draw[->,dotted] (\Td,\Xd) -- (\Td,\Xe);

      \draw[domain=\Td:\Te, line width=0.5mm,smooth,variable=\x,red] plot ({\x},\Xe);
     
\end{tikzpicture}
}
\caption{(Top) Graph for the up-down change-point model proposed  in \cite{hocking2015peakseg}. We have $\mathcal{S}=\{Up, Dw\}$, the loss function is always the $\ell_2$. The penalty is omitted when equal to zero. (Bottom) In red, a piecewise constant function validating the graph of constraints.
The penalty is omitted when equal to zero.}
\label{fig:updown_col}
\end{figure}
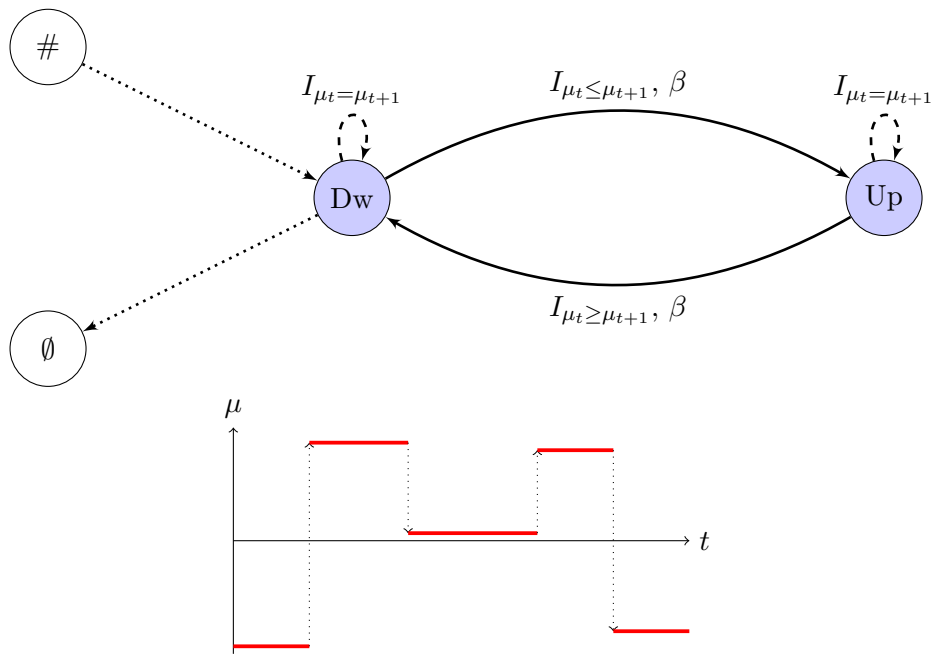

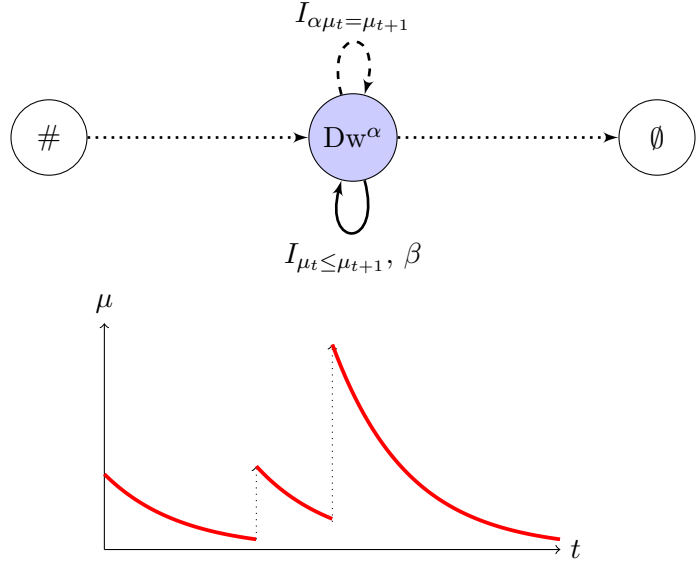
\begin{figure}[!htbp]
\centering{
\begin{tikzpicture}[>=latex']
 \node[draw,circle, minimum size=1cm] at (0, 0) (step_0) {$\#$};
 \node[draw,circle, minimum size=1cm] at (8, 0) (step_{n+1}) {$\emptyset$};
 \node[draw,circle,fill=blue!20, minimum size=1cm] at (4, 0) (state_Dw) {Dw$^\alpha$};
 
 \draw [->, dotted, line width = 1pt] (step_0) to 
		    (state_Dw) ;
 
 \draw [->, dashed, loop above, line width = 1pt] (state_Dw) to 
			node[midway, above=0]{$I_{\alpha \mu_t = \mu_{t+1}}$} 
		    (state_Dw) ;
\draw [->, loop below, line width = 1pt] (state_Dw) to 
			node[midway, below=0]{$I_{\mu_t \leq \mu_{t+1}}$, $\beta$} 
		    (state_Dw) ;
\draw [->, dotted, line width = 1pt] (state_Dw) to (step_{n+1}) ;
 
\end{tikzpicture}
\begin{tikzpicture}
      \draw[->] (0,0) -- (6,0) node[right] {$t$};
      \draw[->] (0,0) -- (0,3) node[above] {$\mu$};

     \draw[samples=400,domain=0:2, line width=0.5mm,smooth,variable=\x,red] plot ({\x}, {\fpeval{exp(-\x)}});
     \draw[->,dotted] (2,{\fpeval{exp(-2)}}) -- (2,{\fpeval{exp(0.1)}});

     \draw[samples=400,domain=2:3, line width=0.5mm,smooth,variable=\x,red] plot ({\x}, {\fpeval{exp(-\x+2.1)}});
     \draw[->,dotted] (3,{\fpeval{exp(-1)}}) -- (3,{\fpeval{exp(1)}});

     \draw[samples=400,domain=3:6, line width=0.5mm,smooth,variable=\x,red] plot ({\x}, {\fpeval{exp(-\x+4)}});
     
\end{tikzpicture}
}
\caption{(Top) Graph for the up - exponential decrease change-point model proposed in \cite{jewell2018fast}. We have $\mathcal{S}=\{Dw^\alpha\}$, the loss function is always the $\ell_2$. The penalty is omitted when equal to zero. (Bottom) In red, a function validating the graph of constraints.}
\label{fig:updownexp_col}
\end{figure}

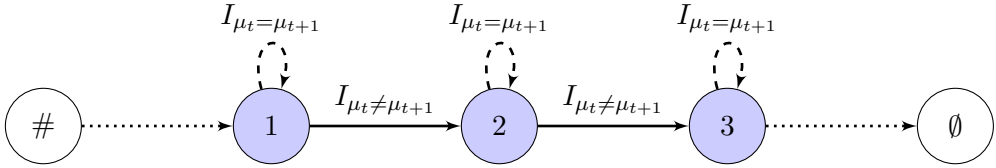
\begin{figure}[!htbp]
\centering{
\begin{tikzpicture}[>=latex']
 \node[draw,circle, minimum size=1cm] at (0, 0) (step_0) {$\#$};
 \node[draw,circle, minimum size=1cm] at (12, 0) (step_{n+1}) {$\emptyset$};
 \node[draw,circle,fill=blue!20, minimum size=1cm] at (3, 0) (state_1) {1};
 \node[draw,circle,fill=blue!20, minimum size=1cm] at (6, 0) (state_2) {2};
 \node[draw,circle,fill=blue!20, minimum size=1cm] at (9, 0) (state_3) {3};

  \draw [->, dotted, line width = 1pt] (step_0) to (state_1) ;
 \draw [->, dashed, loop above, line width = 1pt] (state_1) to 
			node[midway, above=0]{$I_{\mu_t = \mu_{t+1}}$} 
		    (state_1) ;
 \draw [->, line width = 1pt] (state_1) to 
			node[midway, above=0]{$I_{\mu_t \neq \mu_{t+1}}$} 
		    (state_2) ;

 \draw [->, dashed, loop above, line width = 1pt] (state_2) to 
			node[midway, above=0]{$I_{\mu_t = \mu_{t+1}}$} 
		    (state_2) ;
 \draw [->, line width = 1pt] (state_2) to 
			node[midway, above=0]{$I_{\mu_t \neq \mu_{t+1}}$} 
		    (state_3) ;
 \draw [->, dashed, line width = 1pt, loop above] (state_3) to 
			node[midway, above=0]{$I_{\mu_t = \mu_{t+1}}$} 
		    (state_3) ;
 \draw [->, dotted, line width = 1pt] (state_3) to (step_{n+1}) ;

\end{tikzpicture}
}
\caption{Graph for the 3-segment change-point model. We have $\mathcal{S}=\{1, 2, 3\}$, the loss function is always the $\ell_2$. The penalty is always equal to zero.}
\label{fig:3seg}
\end{figure}

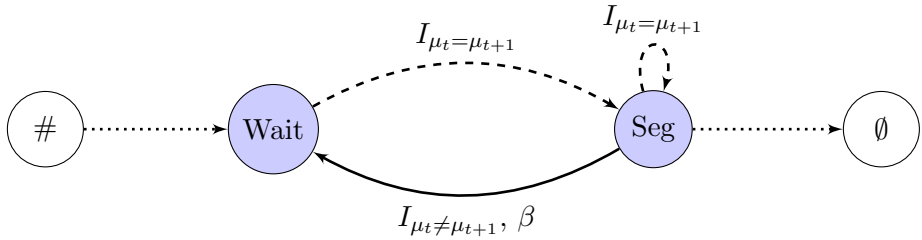
\begin{figure}[!htbp]
\centering{
\begin{tikzpicture}[>=latex']
 \node[draw,circle, minimum size=1cm] at (0, 0) (step_0) {$\#$};
 \node[draw,circle,, minimum size=1cm] at (11, 0) (step_{n+1}) {$\emptyset$};
 \node[draw,circle,fill=blue!20, minimum size=1cm] at (8, 0) (state_Seg) {Seg};
 \node[draw,circle,fill=blue!20, minimum size=1cm] at (3, 0) (state_Wait) {Wait};
 
 \draw [->, dotted, line width = 1pt] (step_0) to (state_Wait) ;
 \draw [->, dashed, loop above, line width = 1pt] (state_Seg) to 
			node[midway, above=0]{$I_{\mu_t = \mu_{t+1}}$} 
		    (state_Seg) ;
 \draw [->, bend left, line width = 1pt] (state_Seg) to 
			node[midway, below=0] {$I_{\mu_t \neq \mu_{t+1}}$, $\beta$} 
		    (state_Wait) ;
 \draw [->, dotted, line width = 1pt] (state_Seg) to (step_{n+1}) ;
 
 \draw [->, dashed, bend left, line width = 1pt] (state_Wait) to
			node[midway, above=0]{$I_{\mu_t = \mu_{t+1}}$} 
		    (state_Seg) ;
\end{tikzpicture}
}
\caption{Graph for the at least 2 data-points per segment change-point model. We have $\mathcal{S}=\{Wait, Seg\}$,
the loss function is always the $\ell_2$. The penalty is omitted when equal to zero.}
\label{fig:atleast2}
\end{figure}

\begin{itemize}
\item (Up - Down) To model peaks \cite{hocking2015peakseg} proposed an up-down constraint using two states $\mathcal{S} = \{Up, Dw\}$. Transitions from $Dw$ to $Up$ are forced to go up $I_{\mu_t \leq \mu_{t+1}}$. Transitions from $Up$ to $Dw$ are forced to go down $I_{\mu_t \geq \mu_{t+1}}$.  The graph of this model is given in Figure~\ref{fig:updown_col}.

\item(Up - Exponentially Down) To model pulses \cite{jewell2018fast} proposed a model where the mean decreases exponentially between positive spikes. In that case a unique state with two transitions is sufficient. 
The first transition corresponds to an up change $I_{\mu_t \leq \mu_{t+1}}$ and the second to an exponential decay $I_{\alpha \mu_t = \mu_{t+1}}$ with $0 < \alpha < 1$. The graph of this model is given in Figure~\ref{fig:updownexp_col}.

\item (Segment Neighborhood) One often considers a known number of segments, $D$ say  \citep{auger1989algorithms}. This is encoded by a graph with $D$ states, $\mathcal{S}=\{1, ... D\}$. From any $d \in \mathcal{S}$ there are two transitions to consider. One from $d$ to $d$ with constraint $I_{\mu_t=\mu_{t+1}}$ and one from $d$ to $d+1$ with constraints $I_{\mu_t \neq \mu_{t+1}}$. The graph of this model for $D=3$ is given in Figure~\ref{fig:3seg}.

\item (At least 2 data-points per segment) It is often desirable to impose a minimum segment length. For at least 2 data-points one should consider two states $\mathcal{S}=\{Wait, Seg\}$. There are 3 transitions to consider
one from $Seg$ to $Wait$ with the constraint $I_{\mu_t \neq \mu_{t+1}}$, one from $Seg$ to $Seg$ with $I_{\mu_t=\mu_{t+1}}$ and one from $Wait$ to $Seg$ with $I_{\mu_t=\mu_{t+1}}$. The graph of this model is given in Figure~\ref{fig:atleast2}. This can be extended to $p$ data-points. The graph for at least 3 data-points per segment is given in Appendix  \ref{app-sec:graphs} (Figure~\ref{fig:atleast3}).

In Appendix \ref{app-sec:graphs} we  provide a few more examples. In particular, we reformulate the collective anomaly model of \cite{fisch2018linear} as a constrained model.

\end{itemize}


\newpage
\section[Optimization problem solved by gfpop]{Optimization problem solved by \pkg{gfpop}}
\label{sec:optim}

\subsection{Penalized maximum likelihood}\label{sec:prob}
We now present the constrained change-point optimization problem. 
The goal is to minimize the negative log-likelihood over all model parameters that validate the constraints (see Section \ref{subsec:collapsed}):
\begin{displaymath}\label{eq:problem_1}
Q_n = \underset{ \underset{\mu | p(\mu)}{p  = (v, e) \,\in\, \mathcal{G}_n} }{\min} 
\left\{ \sum_{t=1}^n (\gamma_{e_t}(y_t, \mu_t) + \beta_{e_t}) \right\}\,.
\end{displaymath}
This is a discrete optimization problem. A naive exploration of the $2^{n-1}$ 
change-point positions is not feasible in practice. Due to the
constraints, segments are dependent and $Q_n$ cannot be written as a sum over all
segments. Therefore the algorithms of \cite{auger1989algorithms,jackson2005algorithm,killick2012optimal} are not applicable.

\cite{hocking2020constrained} have shown that it is possible to optimize $Q_n$ using functional dynamic programming techniques. The idea is to consider the quantity $Q_n$ as a function of the mean and the state of the last data-point:

\begin{equation}\label{eq:func_cost_opt}
Q_n^s(\theta) = \underset{ \underset{\underset{\mu_n = \theta\,,\, v_n =(n,s)}{\mu \,|\, p(\mu)}}{ p = (v, e) \,\in\, \mathcal{G}_n}}{\min} 
\left\{ \sum_{t=1}^n (\gamma_{e_{t}}(y_t, \mu_t) + \beta_{e_t}) \right\},
\end{equation}
where we use the subscript $n$ to denote the number of data points analyzed, $s$ to denote the state of the most recent transition, and $\theta$ the mean of the last data point.

By construction, each $Q_n^s$ is a piecewise function and can be defined as the pointwise minimum of a finite number of functions, with the form of these functions depending on the loss used. In the package three analytical decompositions for the pieces of $Q_n^s$ are implemented:
\begin{description}
\item[L2 decomposition.] $f_1: \theta \mapsto 1$, $f_2: \theta \mapsto \theta$ and $f_3: \theta \mapsto \theta^2$. This decomposition allows one to consider Gaussian (least-square), biweight and Huber loss functions; 

\item[Lin-log decomposition.] $f_1: \theta \mapsto 1$, $f_2: \theta \mapsto \theta$ and $f_3: \theta \mapsto \log(\theta)$. This decomposition allows one to consider loss functions for Poisson and exponential models. It is also possible to consider a change in the variance of a Gaussian distribution of mean $0$\footnote{To be clear, in that case the log-likelihood is $\frac{1}{2} \log(\frac{1}{\sigma^2}) -\frac{y_t}{2\sigma^2}$ and we get the Lin-Log decomposition by taking $\theta=\frac{1}{\sigma^2}$};

\item[Log-log decomposition.] $f_1: \theta \mapsto 1$, $f_2: \theta \mapsto \log(\theta)$ and $f_3: \theta \mapsto \log(1-\theta)$. This decomposition allows one to consider loss functions for the binomial and negative binomial likelihoods.
\end{description}
 
As in the Viterbi algorithm for finite state space HMM, it is possible to define an update formula linking the set of functions $\{\theta  \mapsto Q_{n-1}^{s}(\theta)\,,\, s \in \mathcal{S} \}$ to $\theta \mapsto Q_n^{s'} (\theta)$ for all states $s'$. Computationally, the update is applied per interval using some edge-dependent operators described in the following subsection \citep{rigaill2015pruned,maidstone2017optimal,hocking2020constrained}.

\subsection[Operators and update-rule for gfpop]{Operators and update-rule for \code{gfpop}}\label{sec:opandupdate}

Let us consider a transition from $s$ to $s'$ at step $n$.
Its edge is $(s, s')$ and its associated constraint is $I_{(s, s')}$. The \code{gfpop} algorithm involves calculating the best $(\theta, s)$ to reach state $(\theta', s')$, i.e., minimizing the functional cost while satisfying the constraint $I_{(s, s')}$. Formally this is defined as an operator:
\begin{eqnarray*}
O_n^{s,s'}(\theta') = \underset{\theta | I_{(s, s')}(\theta,\theta') }{\min} \{ Q_n^{s} (\theta) \}\,.
\end{eqnarray*}

\paragraph{Operator calculation}

For a general constraint $I$ and a general function $Q_n^s$ it is not easy to compute $O_n^{s, s'}$. Recall that $Q_n^{s}(\theta)$ are piecewise analytical, i.e., they can be exactly represented by a finite set of real-valued coefficients. For algorithmic simplicity \cite{hocking2020constrained} requires that $O_n^{s,s'}(\theta)$ has
the same analytical decomposition per interval (L2, Lin-Log or Log-Log).

In practice here are the constraints we can accommodate. 
\begin{description}
\item[L2 decomposition:] any linear constraint, e.g., $a \mu_t + b \mu_{t+1} + c  \leq 0$ or $a \mu_t + b \mu_{t+1} + c  = 0$;
\item[Lin-log decomposition:] any proportional constraint e.g., $a \mu_t \leq \mu_{t+1}$ or $a \mu_t = \mu_{t+1}$; 
\item[Log-log decomposition:] only the two inequalities $\mu_t \leq \mu_{t+1}$ or $\mu_t \geq \mu_{t+1}$.
\end{description}

Note that constraints can be combined by considering more than one edge from one state to another. In particular, for L2 decomposition the constraint $|\mu_{t+1}-\mu_{t}| \geq c$ can be implemented using $\mu_t + c \leq \mu_{t+1}$ or $\mu_t \geq \mu_{t+1} + c$. This constraint encodes the idea of detecting sufficiently large changes (also called relevant changes) described in \cite{dette2016detecting}. 

Computationally, it is possible to compute $O_n^{s,s'}(\theta)$ by scanning from left to right or from right to left all intervals which correspond to a different functional form of $Q_n^{s} (\theta)$ (See examples in \cite{hocking2020constrained}).

\paragraph{Update-rule.}
Given this operator function we can now define the update rule used by the \code{gfpop} algorithm.
\begin{eqnarray}\label{eq:update_rule}
Q_{n+1}^{s'} (\theta) = \min_{s | \exists \,edge \,(s, s')} \left\{ O_n^{s,s'}(\theta) + \gamma_{(s, s')}(y_{n+1},\theta) + \beta_{(s, s')}  \right\}\,.
\end{eqnarray}

For simplicity, we do not describe the update for initial and final steps. The proof of this update rule is very similar to the proof of the Viterbi algorithm and is given in Appendix \ref{app-sec:proof-update}). It follows the strategy of \cite{hocking2020constrained}. Notice also that recovering the optimal set of change-points from all $Q_{1}^{s},...,Q_{n}^{s}$ by backtracking is not straightforward because of the need to validate the constraints between consecutive segments. We provide some details in Appendix \ref{app-sec:backtracking}.

\paragraph{An example with fpop.} With the standard multiple change-point model (see Section \ref{subsec:std}  and Figure~\ref{fig:std_col}) we have only one vertex (1) and two edges denoted here $0$ (no change) and $1$ (a change) replacing the notation $(s,s')$. We get with Equation \ref{eq:update_rule}:
\begin{eqnarray*}
Q_{n+1}^{1} (\theta) = \min \left\{ O_n^0(\theta) + \gamma_{0}(y_{n+1},\theta) + \beta_{0} ,\, O_n^{1}(\theta) + \gamma_{1}(y_{n+1},\theta) + \beta_{1}  \right\}\,.
\end{eqnarray*}
Only edge $1$ is penalized, so that $\beta_0 = 0$ and $\beta_1 = \beta > 0$. As we have no robust loss or parameter constraint on the cost function: $\gamma_{0}(\cdot,\cdot) = \gamma_{1}(\cdot,\cdot) = \gamma(\cdot,\cdot)$. For edge $0$, $O_n^0(\theta') = \underset{\theta | \theta =\theta'}{\min} \{ Q_n^{1} (\theta) \} = Q_n^{1} (\theta')$ and for edge $1$, $O_n^1(\theta') = \underset{\theta | \theta \ne \theta'}{\min} \{ Q_n^{1} (\theta) \} =  \underset{\theta}{\min} \{ Q_n^{1} (\theta) \}$. Removing the state index, we eventually obtain the well-known FPOP update-rule:
\begin{eqnarray*}
Q_{n+1}(\theta) = \min \left\{Q_n(\theta),\, \underset{\theta}{\min} \{ Q_n (\theta)\} + \beta  \right\} + \gamma(y_{n+1},\theta)\,.
\end{eqnarray*}
If we assume a Gaussian loss for change in mean, we have $\gamma(y_{n+1}, \theta) = (y_{n+1} - \theta)^2$, quadratic in $\theta$. The update consists in reconstructing the optimal cost by finding for all $\theta$ the minimum between $Q_{n}(\theta)$ and the $\underset{\theta}{\min} \{ Q_n (\theta)\}$ constant line leading to a function $Q_{n+1}(\cdot)$ piecewise quadratic in $\theta$. Ways to deal efficiently with this update rule have been presented in \cite{maidstone2017optimal}. For implementing the constraints included in the package, see \cite{hocking2020constrained} and \cite{JSSv101i10}; for other loss functions see also \cite{fearnhead2018changepoint} and \cite{jewell2018fast}.

\subsection{How to choose the loss function and penalty}

The choice of the loss function $\gamma$ is linked to the choice of the noise model.
This choice is not necessarily easy. For example for continuous data
it might make sense to consider the least square error \citep{picard2005statistical}; in the presence of outliers considering a robust loss is natural \citep{fearnhead2018changepoint};
and for count data a Poisson loss is often used \citep{hocking2020constrained}.
It is our experience that visualizing the data beforehand is a good way to avoid simple modeling mistakes.
~\\

The choice of the penalty $\beta$ is critical to select the number of change-points.
In the absence of constraint several penalties have been proposed.
For detecting a change in mean with independent Gaussian data, a penalty of $\beta = 2\sigma^2 \log(n)$ was proposed by \cite{yao1989least}. It tends to work well when the number of changes is small. More complex penalties exist, e.g., \cite{zhang2007modified,lebarbier2005detecting,baraud2009gaussian}. For penalties that are concave in the number of segments one can run the Operators and update-rule for \fct{gfpop} algorithm for various values of $\beta$ and recover several segmentations (with a varying number of change-points) \citep{killick2012optimal}. This can be done efficiently using the CROPS algorithm \citep{haynes2014efficient}. In labeled data sets, supervised learning algorithms can be used to infer an accurate model for predicting penalty values $\beta$  \citep{HOCKING-icml-2013,hocking2015peakseg,hocking2020constrained}.
~\\

For models with constraints, to the best of our knowledge there is very little statistical literature available. The paper of \cite{gao2017estimation} describes a penalty in the isotonic case but it was not calibrated. 
It is our experience that the penalties proposed for the unconstrained case tend to work reasonably well, although they are probably sub-optimal from a statistical perspective.

\section[The gfpop package]{The \pkg{gfpop} package}
\label{sec:howto}

\subsection{Graph construction}\label{graphconstruction}

Our \pkg{gfpop} package deals with collapsed graphs for which all the cost functions $\gamma$ have the same decomposition (L2, Lin-log or Log-log). All other characteristics are local and fixed per edge. The graph $\mathcal{G}_n$ (see Section \ref{subsec:collapsed}) is defined in the \pkg{gfpop} package by a collection of edges.

\paragraph{Edge parameters.} An edge is a list of four main elements:
\begin{itemize}
    \item \code{state1}: the starting node defined by a string;
    \item \code{state2}: the arrival node defined by a string;
    \item \code{type}: a string equal to \code{null}, \code{std}, \code{up}, \code{down} or \code{abs} defining the type of constraints between successive nodes respectively corresponding to indicators $I_{\mu_t=\mu_{t+1}}$, $I_{\mu_t \ne \mu_{t+1}}$, $I_{\mu_t + c \le \mu_{t+1}}$, $I_{\mu_t\ge \mu_{t+1} + c}$ and $I_{|\mu_{t+1}-\mu_t| \ge c}$;
    \item \code{penalty}: the penalty $\beta_e$ associated to this edge (it can be zero);
\end{itemize}
and some optional elements:
\begin{itemize}
    \item \code{decay}: a number between $0$ and $1$ for the mean exponential decay (in case type is \code{null}) corresponding to the constraint $I_{\mu_{t+1} =  \alpha\mu_t}$;
    \item gap: the gap $c$ between successive means of the \code{up}, \code{down} and \code{abs} types;
    \item \code{K}: the threshold for the biweight and Huber losses ($K > 0$);
    \item \code{a}: the slope for the Huber robust loss ($a \ge 0$).
\end{itemize}

\paragraph{An example of an edge.} We can define an edge \code{e1} with the function \code{Edge} as:

\begin{CodeChunk}
\begin{CodeInput}
R> e1 <- Edge(state1 = "Dw", state2 = "Up",
                            type = "up", penalty = 10, gap = 0.5)
\end{CodeInput}
\end{CodeChunk}

which is an edge from node \code{Dw} to node \code{Up} with an up constraint, penalty $\beta = 10$ and a minimal jump size of $0.5$.

\paragraph{An example of a graph.} We provide an example of graph for collective anomalies detection with the \pkg{gfpop} package given in Figure~\ref{fig:collective} (see \cite{fisch2018linear}):

\begin{CodeChunk}
\begin{CodeInput}
R> graph(
+     Edge(state1 = "mu0", state2 = "mu0", penalty = 0, K = 3),
+     Edge(state1 = "mu0", state2 = "Coll", penalty = 10, type = "std"),
+     Edge(state1 = "Coll", state2 = "Coll", penalty = 0),
+     Edge(state1 = "Coll", state2 = "mu0", penalty = 0, type = "std", K = 3), 
+     StartEnd(start = "mu0", end = c("mu0", "Coll")),
+     Node(state = "mu0", min = 0, max = 0)
+     )
\end{CodeInput}
\begin{CodeOutput}
  state1 state2  type parameter penalty   K  a min max
1    mu0    mu0  null         1       0   3  0  NA  NA
2    mu0   Coll   std         0      10 Inf  0  NA  NA
3   Coll   Coll  null         1       0 Inf  0  NA  NA
4   Coll    mu0   std         0       0   3  0  NA  NA
5    mu0   <NA> start        NA      NA  NA NA  NA  NA
6    mu0   <NA>   end        NA      NA  NA NA  NA  NA
7   Coll   <NA>   end        NA      NA  NA NA  NA  NA
8    mu0    mu0  node        NA      NA  NA NA   0   0
\end{CodeOutput}
\end{CodeChunk}

Notice that the graph is encoded into a data-frame.

\paragraph{Note 1.} Most graphs (such as the previous one) contain recursive edges, that is edges with the same starting and arrival node. The absence of this edge forces a change and is useful to enforce a minimal segment length (see Figures~\ref{fig:atleast2} and \ref{fig:atleast3}). 

\paragraph{Note 2.} 
In the \pkg{gfpop} graph definition, a starting (resp. arrival) node is a state $s$ for which there exists an edge between the starting $v_0 = (0,\#)$ (resp. arrival $v_{n+1} = (n+1,\emptyset)$) node and $s$ (See Section \ref{subsec:collapsed}). These specific states are defined using function \code{StartEnd}. If not specified, all nodes are starting and arrival nodes. The range of values for parameter inference at each node can be constrained using function \code{Node}. 

In this example we have two states, \code{mu0} and \code{coll}. Both states can be arrival states, but we have fixed the start node to be \code{mu0}. This node \code{mu0} is restricted by \code{min = 0} and \code{max = 0} using the \code{Node} function, such that only the zero value can be inferred for any segment in that state.

\paragraph{Some default graphs.} We included in function \fct{graph} the possibility to directly build some standard graphs. Here is an example for the isotonic case corresponding to Figure~\ref{fig:iso_col}:

\begin{CodeChunk}
\begin{CodeInput}
R> graph(type = "isotonic", penalty = 12)
\end{CodeInput}
\begin{CodeOutput}
  state1 state2 type parameter penalty   K a min max
1    Iso    Iso null         1       0 Inf 0  NA  NA
2    Iso    Iso   up         0      12 Inf 0  NA  NA
\end{CodeOutput}
\end{CodeChunk}

Three other standard graph types are: \code{std}, \code{updown} corresponding to Figures~\ref{fig:std_col} and \ref{fig:updown_col} and \code{relevant} corresponding to Figure~\ref{fig:relevant}. All graphs presented in this paper are available in our package through the function \fct{paperGraph}, where its first parameter is the figure number.

\subsection[The gfpop function]{The \fct{gfpop} function}

The \fct{gfpop} function takes as an input the data and the graph and runs the algorithm. It returns a set of change-points and the non-penalized cost (that is the value of the fit to the data ignoring the penalties for adding changes). It also returns the mean value and the state of each segment. The boolean \code{forced} value indicates whether a linear inequality constraint is active, which means that the $\mu_t$ and $\mu_{t+1}$ values lie on the frontier defined by the inequality constraint. Below we illustrate the use of the \fct{gfpop} function for various graphs and loss functions.

We first simulate data. To do this we use the \fct{dataGenerator} function provided by the \pkg{gfpop} package. The function generate \code{n} data-points using a distribution of \code{type} \code{mean} (by default), \code{poisson}, \code{exp}, \code{variance} or \code{negbin} following a change-point model given by relative change-point positions (a vector of increasing values in $(0,1]$). Standard deviation parameter \code{sigma} and decay \code{gamma} are specific to the Gaussian mean model, whereas  \code{size} is linked to the \proglang{R} \code{rnbinom} function from \proglang{R} \code{stats} package.

\paragraph{Gaussian model with an up-down graph.}

Here is an example with a Gaussian cost and a standard penalty of $2 \log(n)$ for the up-down graph. We simulate data from a change in mean model with Gaussian observations.
\begin{CodeChunk}
\begin{CodeInput}
R> set.seed(75)
R> n <- 1000
R> myData <- dataGenerator(n, c(0.1, 0.3, 0.5, 0.8, 1), 
+                           c(1, 2, 1, 3, 1), sigma = 1)
\end{CodeInput}
\end{CodeChunk}
This data has $5$ segments, with the end of segments at relative positions $0.1$, $0.3$, $0.5$, $0.8$ and $1$ along the $n = 1000$ data points; and with segment means being respectively $1, 2, 1, 3$ and $1$.
\begin{CodeChunk}
\begin{CodeInput}
R> set.seed(75)
R> n <- 1000
R> myData <- dataGenerator(n, c(0.1, 0.3, 0.5, 0.8, 1), 
+                           c(1, 2, 1, 3, 1), sigma = 1)
R> myGraph <- graph(penalty = 2 * log(n), type = "updown")
R> gfpop(data = myData, mygraph = myGraph, type = "mean")
\end{CodeInput}
\begin{CodeOutput}
$changepoints
[1]  108  295  500  800 1000

$states
[1] "Dw" "Up" "Dw" "Up" "Dw"

$forced
[1] FALSE FALSE FALSE FALSE

$parameters
[1] 1.044920 2.047202 1.017550 2.916826 1.030938

$globalCost
[1] 963.0278

attr(,"class")
[1] "gfpop" "mean"
\end{CodeOutput}
\end{CodeChunk}
The call to the \code{gfpop} function requires specifying the data, the graph that encapsulates the changepoint model, and the type of loss function. Here \code{type = "mean"} specifies the use of the L2 or biweight loss.

The response contains four vectors. A vector \code{changepoints} contains the last index of each segment, a vector \code{states} gives the nodes in which lie the successive parameter values of the \code{parameters} vector. The vector \code{forced} is a vector of booleans of size 'number of segments - 1' with entry \code{TRUE} when the transition between two states (nodes) has been forced. The \code{globalCost} is the non-penalized cost.

\paragraph{Gaussian Robust biweight model with an up-down graph.}
\label{subsec:robust-loss}

Below we illustrate the use of the biweight loss on data where $10\%$ of the data points are outliers. We shift these data by $\pm 5$ using function \code{rbinom} (4th line of code below). We use the biweight loss with $K = 3$ and an \code{updown} graph with a difference of at least $1$ between consecutive means.
\begin{CodeChunk}
\begin{CodeInput}
R> n <- 1000
R> chgtpt <- c(0.1, 0.3, 0.5, 0.8, 1)
R> myData <- dataGenerator(n, chgtpt, c(0, 1, 0, 1, 0), sigma = 1)
R> myData <- myData + 5 * rbinom(n, 1, 0.05) - 5 * rbinom(n, 1, 0.05)
R> beta <- 2 * log(n)
R> myGraph <- graph(
+           Edge("Dw", "Up", type = "up", penalty = beta, gap = 1, K = 3),
+           Edge("Up", "Dw", type = "down", penalty = beta, gap = 1, K = 3),
+           Edge("Dw", "Dw", type = "null", K = 3),
+           Edge("Up", "Up", type = "null", K = 3),
+           StartEnd(start = "Dw", end = "Dw"))
R> gfpop(data =  myData, mygraph = myGraph, type = "mean")
\end{CodeInput}
\begin{CodeOutput}
$changepoints
[1]  102  311  500  806 1000

$states
[1] "Dw" "Up" "Dw" "Up" "Dw"

$forced
[1] TRUE FALSE FALSE FALSE

$parameters
[1] -0.02296768  0.97703232 -0.03434534  1.00246359 -0.03334062

$globalCost
[1] 1097.364

attr(,"class")
[1] "gfpop" "mean" 
\end{CodeOutput}
\end{CodeChunk}
The difference between this graph and the one for the previous example is the specification \code{K = 3} for each edge. This enforces the use of the biweight loss (with $K = 3$) as opposed to the L2 loss.

\paragraph{Poisson model with isotonic up graph.}

We provide an example with a Lin-log cost decomposition with Poisson data constrained to up changes, with the mean at least a doubling at each change.
\begin{CodeChunk}
\begin{CodeInput}
R> n <- 1000
R> chgtpt <- c(0.1, 0.3, 0.5, 0.8, 1)
R> myData <- dataGenerator(n, chgtpt, c(1, 3, 5, 7, 12), type = "poisson")
R> beta <- 2 * log(n)
R> myGraph <- graph(type = "isotonic", gap = 2)
R> gfpop(data =  myData, mygraph = myGraph, type = "poisson")
\end{CodeInput}
\begin{CodeOutput}
$changepoints
[1]    2   99  297  796 1000

$states
[1] "Iso" "Iso" "Iso" "Iso" "Iso"

$forced
[1] TRUE FALSE TRUE TRUE

$parameters
[1]  0.4693878  0.9387755  2.9840954  5.9681909 11.9363817

$globalCost
[1] -5832.845

attr(,"class")
[1] "gfpop"   "poisson"
\end{CodeOutput}
\end{CodeChunk}
The use of Poisson loss is enforced by \code{type = "poisson"} in the call to \code{gfpop}. The graph that describes our change-point model is the default one for isotonic changes, but with the additional constraint on means at least doubling being specified by \code{gap = 2}.

\paragraph{Negative binomial model with 3-segment graph.}

The parameters to find are probabilities and we restrict the inference to 3 segments. The optional parameter \code{all.null.edges} in graph function automatically generates \code{null} edges for all nodes.
\begin{CodeChunk}
\begin{CodeInput}
R> myGraph <- graph(
+   Edge("1", "2", type = "std", penalty = 0),
+   Edge("2", "3", type = "std", penalty = 0),
+   StartEnd(start = "1", end = "3"), 
+   all.null.edges = TRUE)
R> myData <- dataGenerator(n = 1000, changepoints = c(0.3,0.7,1),
+               parameters = c(0.2,0.25,0.3), type = "negbin")
R> gfpop(myData, myGraph, type = "negbin")
\end{CodeInput}
\begin{CodeOutput}
$changepoints
[1]  300  714 1000

$states
[1] "1" "2" "3"

$forced
[1] FALSE FALSE

$parameters
[1] 0.2117808 0.2652162 0.3212748

$globalCost
[1] 2193.216

attr(,"class")
[1] "gfpop"  "negbin"
\end{CodeOutput}
\end{CodeChunk}

Each data-point can be weighted using parameter \code{weights} in \code{gfpop} function. It can be useful to gather consecutive identical values for count data time-series in order to speed-up the change-point analysis \cite{cleynen2014segmentor3isback}.

\subsection[Some additional useful functions in gfpop]{Some additional useful functions in \pkg{gfpop}}

\paragraph{Standard deviation estimation.} For many real-data-sets examples, we are obliged to estimate the standard deviation from the observed data. This value is then used to normalize the data or to be included in edge penalties. The \fct{sdDiff} returns such an estimation with the default HALL method \citep{hall1990asymptotically} well suited for time series with change-points.

\paragraph{A plotting function.} We defined a plotting function \fct{plot}, which shows data-points and the results of the \fct{gfpop} function by using inferred segment parameters and change-points. The user can plot the result in two graphs or only one for \code{mean} and \code{poisson} types (see parameter \code{multiple}) and has to explicitly use the \code{data} parameter as in following examples.

Example 1:
\begin{CodeChunk}
\begin{CodeInput}
R> set.seed(86)
R> myData <- dataGenerator(1000, c(0.3, 0.4, 0.7, 0.95, 1), 
+                      c(1, 3, 1, -1, 4), "mean", sigma = 3)
R> s <- sdDiff(myData)
R> g <- gfpop(myData,
+    graph(type = "relevant", gap = 0.5, penalty = 2 * s ^ 2 * log(1000)),
+    type = "mean")
R> plot(x = g, data = myData, multiple = FALSE)
\end{CodeInput}
\end{CodeChunk}

Example 2:
\begin{CodeChunk}
\begin{CodeInput}
R> set.seed(86) 
R> myData <- dataGenerator(1000, c(0.4, 0.8, 1), c(1, 1.3, 2.3), "exp")
R> s <- sdDiff(myData)
R> g <- gfpop(myData,  type = "exp",
+           graph(type = "isotonic", penalty = 2 * s ^ 2 * log(1000)))
R> plot(x = g, data = myData, multiple = TRUE)
\end{CodeInput}
\end{CodeChunk}

\begin{figure}[ht!]
\centering
\includegraphics[width=0.49\textwidth]{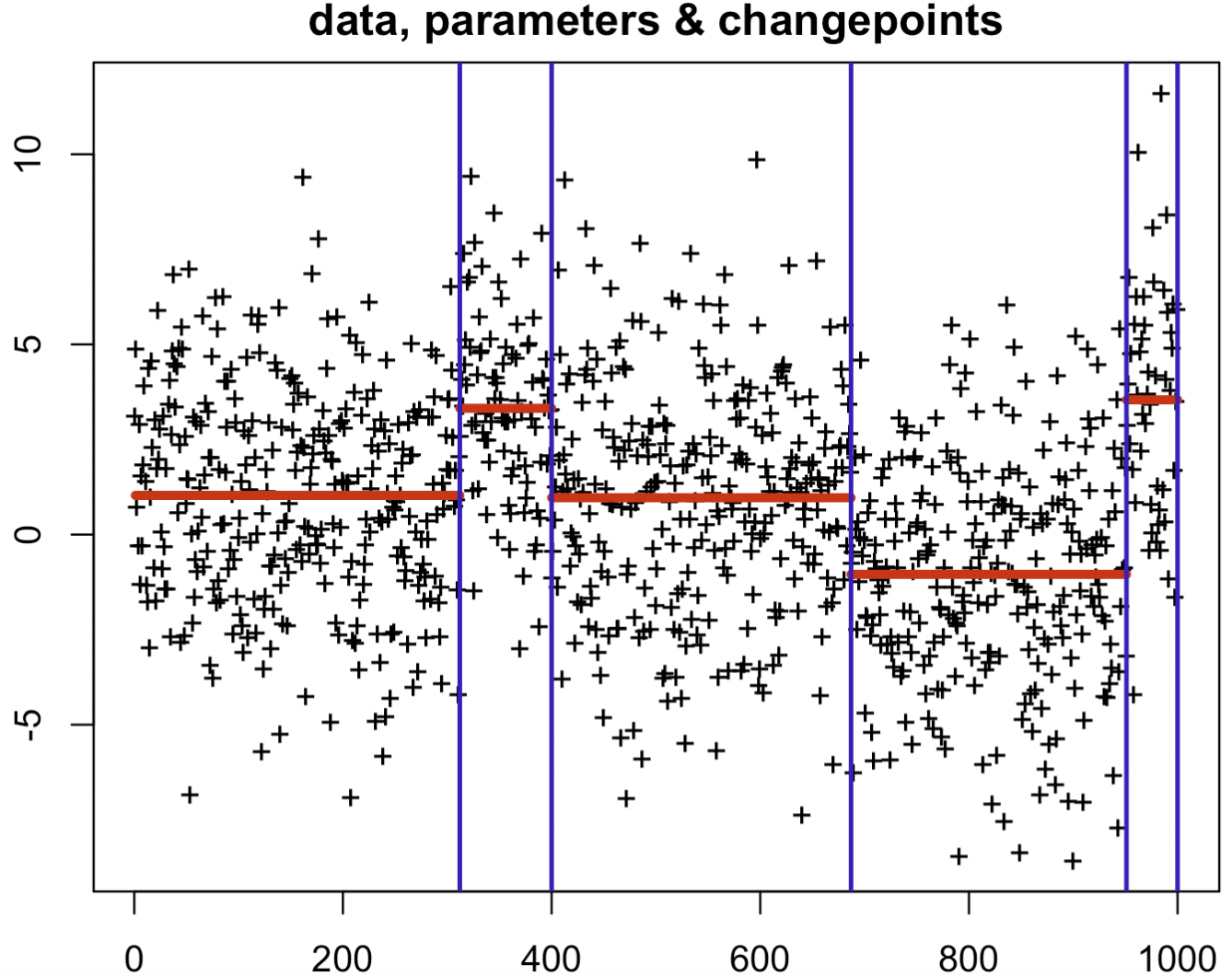}
\includegraphics[width=0.49\textwidth]{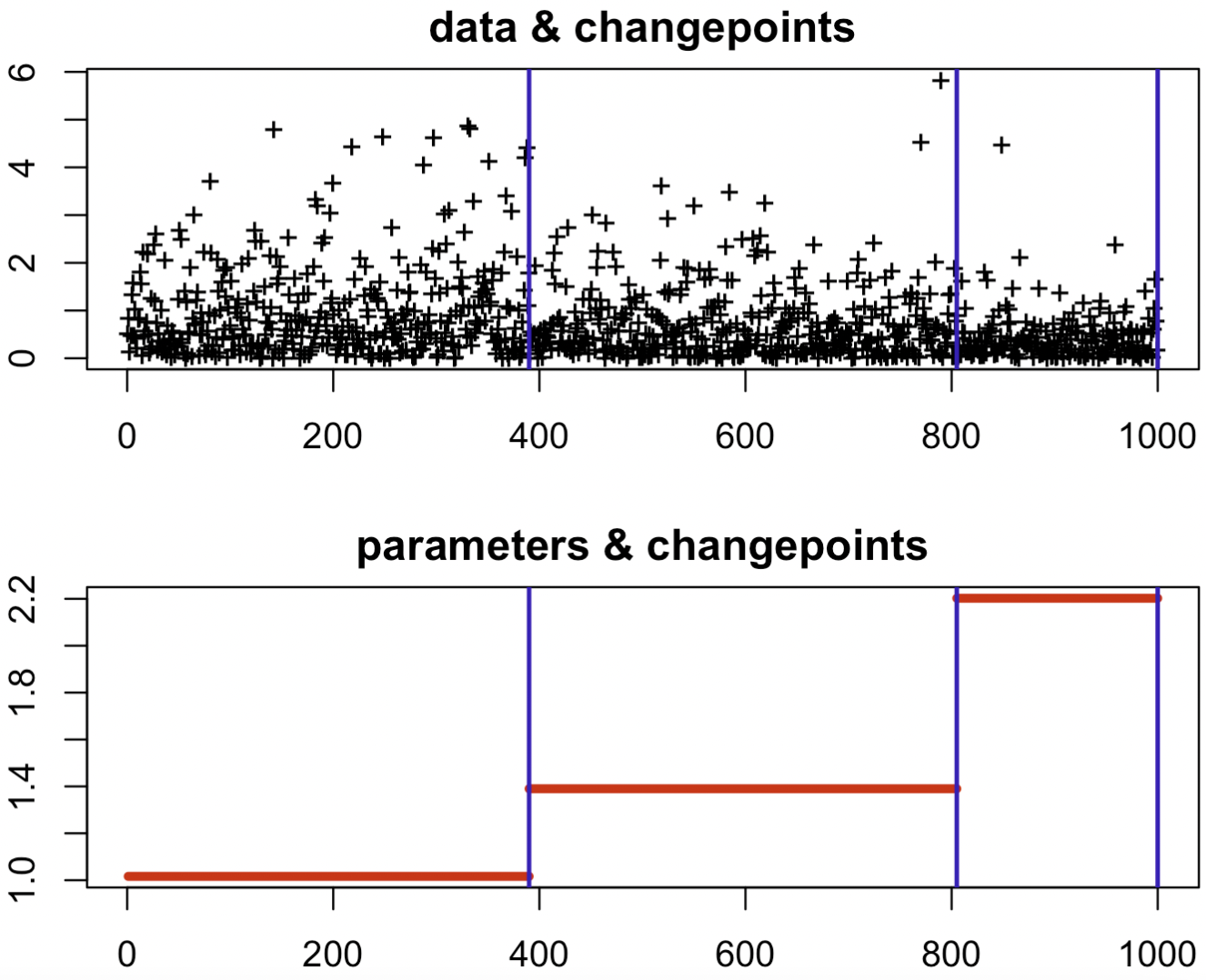}
\caption{The red piecewise constant signal is the $\mu$ vector find by the \fct{gfpop} function, the blue vertical lines indicate the change-point positions. They are built using response vectors \code{changepoints} and \code{parameters}. The left graph presents the result of example~1, the right graph of example~2.}
\end{figure}


\section{Modeling real data with graph-constrained models}
\label{sec:examples}

In this section we illustrate the use of our package on several real datasets. For each application we illustrate several possible sets of constraints and briefly discuss their relative advantages.

\subsection{Gaussian model for DNA copy number data}

We consider DNA copy number data, which are biological measurements that characterize the number of chromosomes in cell samples. Abrupt changes along chromosomes in these data are important indicators of severity in cancers such as neuroblastoma \citep{Schleiermacher2010}.
The non-constrained Gaussian segmentation model has been shown to have state-of-the-art change-point detection accuracy in these data \citep{HOCKING2013}. 

\begin{figure}[!ht]
    \centering
    \begin{tikzpicture}[>=latex']
 \node[draw,circle, minimum size=1cm] at (0, 0) (step_0) {$\#$};
 \node[draw,circle, minimum size=1cm] at (8, 0) (step_{n+1}) {$\emptyset$};
 \node[draw,circle,fill=blue!20, minimum size=1cm] at (4, 0) (state_1) {$1$};
 \draw [->, dotted, line width = 1pt] (step_0) to (state_1) ;
 \draw [->, dashed, loop above, line width = 1pt] (state_1) to 
			node[midway, above=0]{$I_{\mu_t = \mu_{t+1}}$} 
		    (state_1) ;
 \draw [->, loop below=0cm, line width = 1pt] (state_1) to 
			node[midway, below=0]{$I_{|\mu_t - \mu_{t+1}| \geq c }$, $\beta$} 
		    (state_1) ;
 \draw [->, dotted, line width = 1pt] (state_1) to (step_{n+1}) ;

\end{tikzpicture}
    \includegraphics[width=\textwidth]{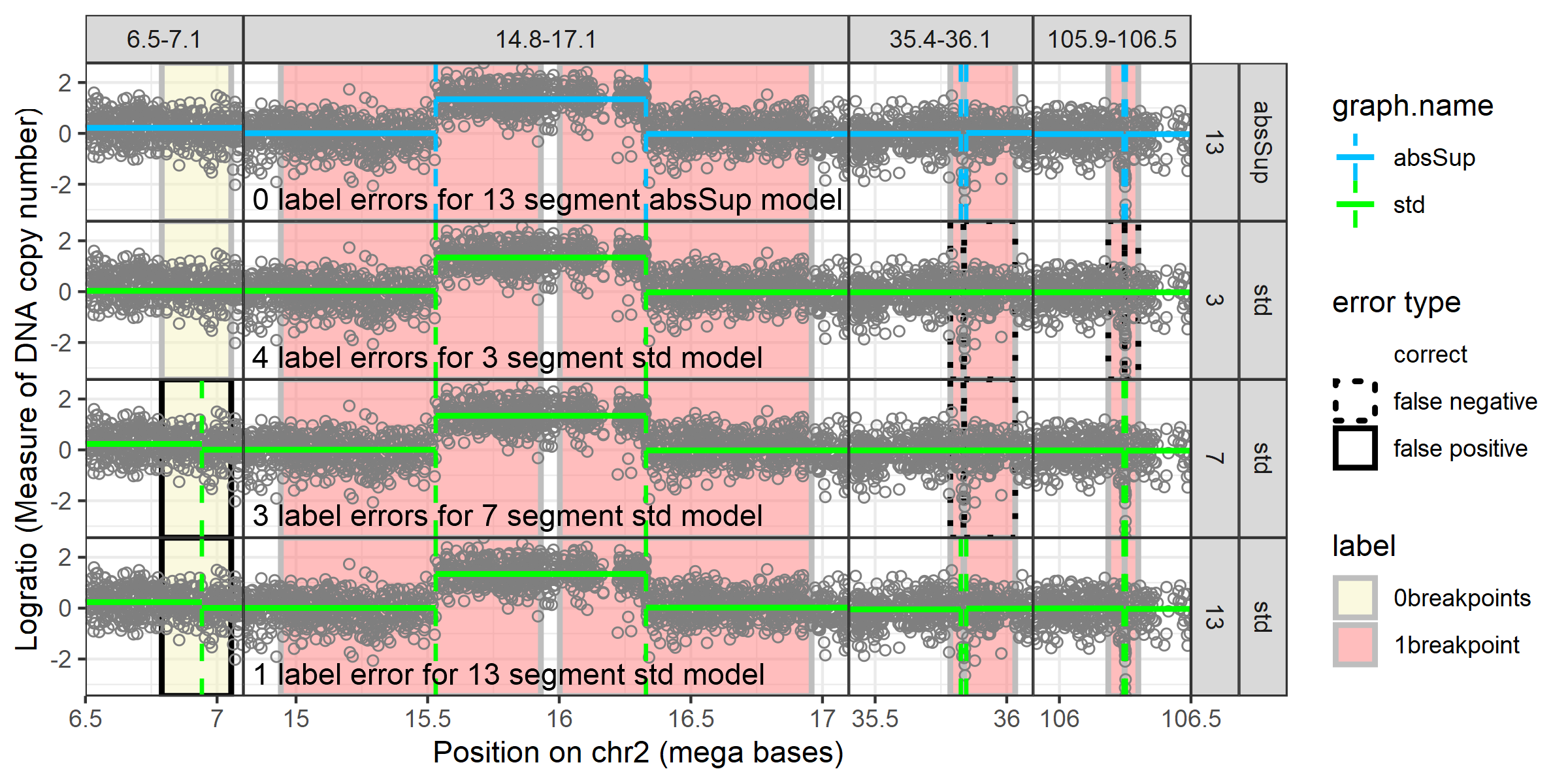}
    \vskip -0.5cm
    \caption{\textbf{Top graph:} Relevant change-point model; all changes are forced to be greater than $a$ in absolute value. 
    \textbf{Below:} Four subsets/windows of a DNA copy number profile (panels from left to right) and four change-point models (panels from top to bottom); rectangles show expert-provided labels which are assumed to be a gold standard. \textbf{Top panel with blue model:} abs model with 13 segments enforces the constraint that the absolute value of each change must be at least 1, $|\mu_{t+1}-\mu_{t}|\geq 1$, which achieves zero label errors in these data. \textbf{Bottom panels with green models:} each model with no constraints between adjacent segment means has label errors (4 false positives for 3 segments, 2 false positives and 1 false negative for 7 segments, 1 false positive for 13 segments).
    }
    \label{fig:relevant-changes-copy-number}
\end{figure}

However, in some high-density copy number data sets, this model incorrectly detects small changes in mean which are not relevant \citep{Hocking-segannot}.
One such data set is shown in Figure~\ref{fig:relevant-changes-copy-number}, which also has positive and negative labels from an expert genomic scientist that indicated regions with (1breakpoint) or without (0breakpoints) relevant change-points. We used these labels to quantify the accuracy of three unconstrained Gaussian change-point models with several different penalties $\beta$.

\begin{itemize}
\item The model with 13 segments predicts a change-point in each positive label (0/6 false negatives), but predicts one change-point in the negative label (1 false positive), for a total of 1 incorrectly predicted label. (bottom panel)
\end{itemize}
\begin{itemize}
\item The model with 7 segments predicts a change-point in four positive labels (2/6 false negatives), and also predicts the false positive change-point in the negative label, for a total of 3 incorrectly predicted labels. (second panel from bottom)
\end{itemize}
\begin{itemize}
\item The model with 3 segments predicts a change-point in only two positive labels (4/6 false negatives), and predicts no change-point in the negative label (0/1 false positive), for a total of 4 incorrectly predicted labels. (second panel from top)
\end{itemize}
We computed all non-constrained Gaussian models from 1 to 20 segments for these data, and none of them were able to provide change-point predictions that perfectly match the expert-provided labels (each model had at least one false positive or false negative). It is thus problematic to use the unconstrained change-point model in this context, because none of the unconstrained models achieve zero label errors.

To solve this problem we propose a graph (Figure~\ref{fig:relevant-changes-copy-number}, top graph) which enforces only ``relevant'' change-points  $|\mu_{t+1}-\mu_{t}|\geq c$, for some relevant threshold $c>0$. For the DNA copy number data set, we set $c=1$ and choose $\beta$ such that the algorithm returns 13 segments (Figure~\ref{fig:relevant-changes-copy-number}, top panel with blue model). 
The proposed model predicts a change-point in each of the positive labels, but does not predict a change-point in the negative label. 
The proposed graph-constrained change-point model is therefore able to predict change-points that perfectly match the expert-provided labels.
If overfitting is a concern with this procedure, we can consider using the two labels in the last region as a test set (105.9-106.5 mega bases), and the other labels as a train set. 
In that case we chose the penalty with minimal errors with respect to the train set labels, and we observed that the test set label error was also minimized.

\subsection{Gaussian multi-modal regression for neuro spike train data}

\begin{figure}[!ht]
    \centering
  \begin{minipage}{1.3in}
    \centering
\begin{tikzpicture}[>=latex']
 \node[draw,circle, minimum size=1cm] at (0, 0) (step_0) {$\#$};
 \node[draw,circle, minimum size=1cm] at (3, 0) (step_{n+1}) {$\emptyset$};
 \node[draw,circle,fill=blue!20, minimum size=1cm] at (1.5, -4) (state_Up) {Up};
 \node[draw,circle,fill=blue!20, minimum size=1cm] at (1.5, -1) (state_Dw) {Dw};
\draw [->, dotted, line width = 1pt] (step_0) to
		    (state_Dw) ;
 \draw [->, bend left,line width = 1pt] (state_Up) to 
			node[sloped, anchor=center, above=0] {$I_{\mu_t\, \geq\, \mu_{t+1}}$} 
		    (state_Dw) ;
\draw [->, loop below,line width = 1pt] (state_Up) to 
			node[midway, below=0]{$I_{\mu_t \,\leq\, \mu_{t+1}}$} 
		    (state_Up) ;
\draw [->, loop above,line width = 1pt] (state_Dw) to 
			node[midway, above=0]{$I_{\mu_t \,\geq \,\mu_{t+1}}$} 
		    (state_Dw) ;
\draw [->, bend left,line width = 1pt] (state_Dw) to 
			node[sloped, anchor = center, above=0]{$I_{\mu_t \,\leq\, \mu_{t+1}}$, $\beta$} 
		    (state_Up) ;
\draw [->, dotted,line width = 1pt] (state_Dw) to (step_{n+1}) ;
\end{tikzpicture}
  \end{minipage}
   \begin{minipage}{4.6in}
   \vskip 0.5cm
    \includegraphics[width=\textwidth]{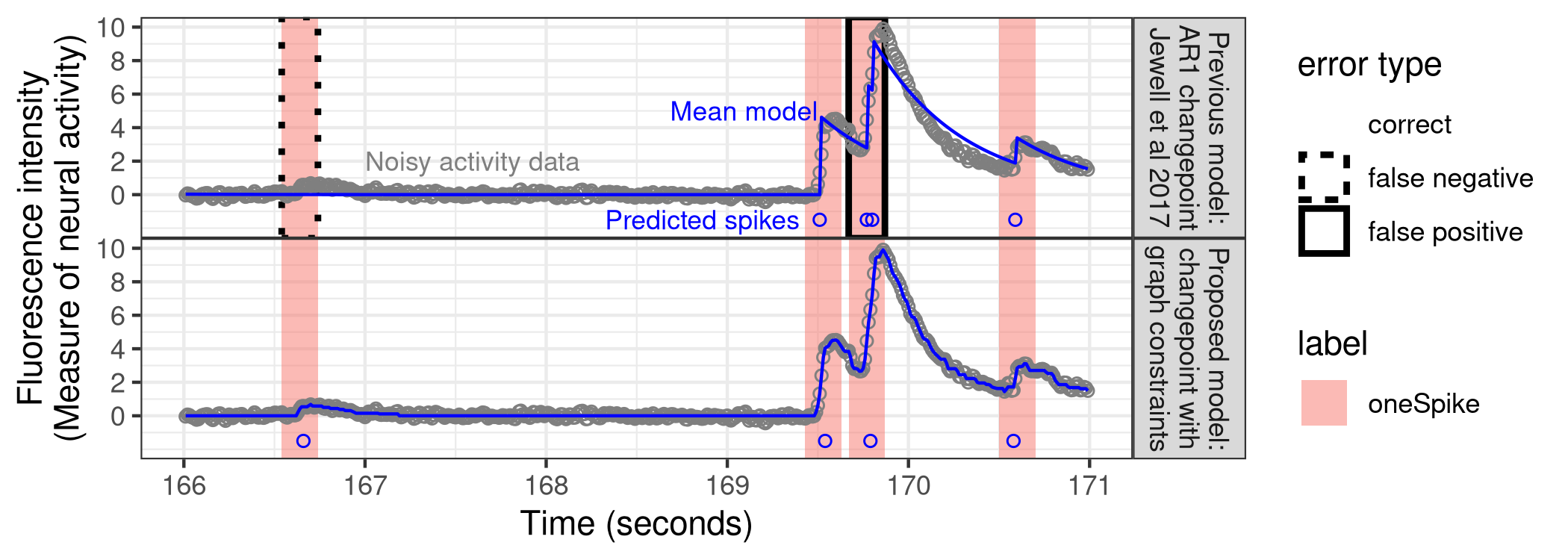}
    \end{minipage}
   
    \caption{\textbf{Left:} Graph for multi-modal regression.  \textbf{Right top:} In these data the previously proposed AR1 model misses a spike in the
    left label (false negative) and predicts two spikes where there
    should be only one in the right label (false
    positive). \textbf{Right bottom:} The proposed 
    multi-modal regression model correctly detects one spike in all the labeled regions.}
    \label{fig:AR1-multimodal}
\end{figure}
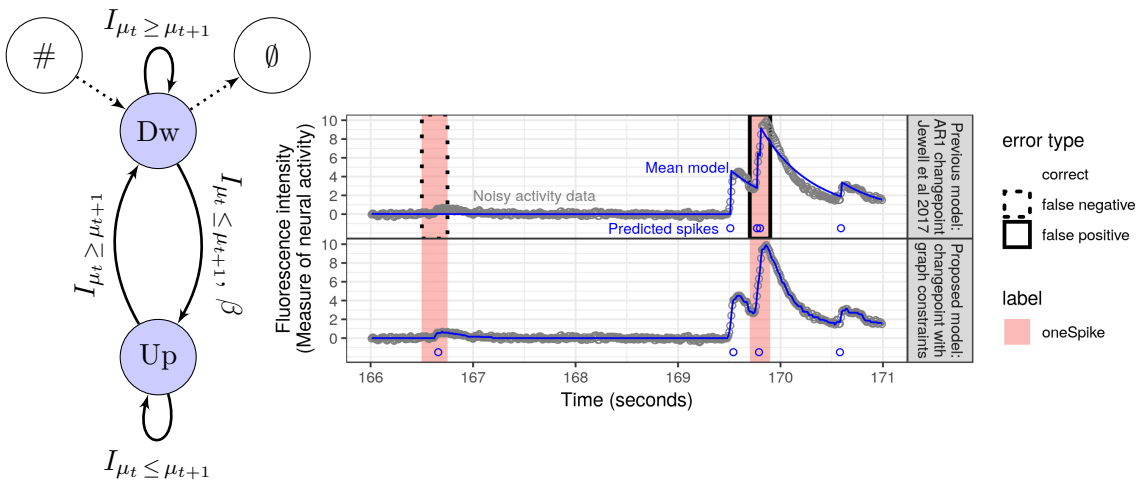

The so-called AR1 change-point model, where the mean decreases exponentially within each segment has been proposed for detecting spikes in calcium imaging data from neuroscience \citep{jewell2018fast}. We fit this model to one calcium imaging data set (Figure~\ref{fig:AR1-multimodal}, Right top) and observed that it is difficult to find a parameter that detects both labeled spikes. 
Red rectangles in Figure~\ref{fig:AR1-multimodal} indicate labels provided by an electrophysiological method which is taken as ground-truth in order to emphasize the qualitative difference between the two algorithms.  
Part of the difficulty of the AR1 model is the fact that there are two parameters to tune, the penalty $\beta$ and also the exponential decay parameter $\gamma$. 
It is more difficult to tune two parameters using grid search because of its quadratic time complexity. Another issues is that a visual inspection of the data suggests that the rate of decay of the mean between spikes may not be constant as assumed by the AR1 model.

We therefore propose a new multi-modal regression model (Isotonic up - Isotonic down graph shown in Figure~\ref{fig:AR1-multimodal}, left) with only one parameter, the penalty $\beta$. We can view this model as detecting modes in the data. Each mode consists of a period before-hand where the mean increases followed by a period where then mean decreases. The period where the mean increases can be interpreted as a period of time where a spike occurs, with the periods where the mean decreases modeling the decay in the data after the spike end. The number of detected spikes is equal to the number of regions where the mean increases, and is controlled by the penalty $\beta$. We observed that it is easy to find a penalty $\beta$ which detects both labeled spikes. Overall these results indicate that the proposed multi-modal regression model (Isotonic up - Isotonic down) is promising for spike detection in calcium imaging data.
We leave a more extensive quantitative comparison to future work.

\subsection{Gaussian nine-state model for electrocardiogram data}

In the context of monitoring hospital patients with heart problems, electrocardiogram (ECG) analysis is one of the most common non-invasive techniques for diagnosing several heart arrhythmia  \citep{Afghah_NIPS}.

A preliminary and fundamental step in ECG analysis is the detection of the QRS complex that leads to detecting the heartbeat and classifying the rhythms \cite{Afghah_ICASSP19}. 

Here, we utilize the proposed change-point detection method to locate the QRS complex in ECG waveforms. The ECG signals used in this study are extracted from the publicly available Physionet Challenge 2015 database ~\cite{PhysioNet,Clifford_Data} that includes measurements for three physiological signals (including ECG) for 750 patients. The resolution and frequency of each signal are 12bit and 250 Hz, respectively. Also, each signal has been filtered by a finite impulse response (FIR) band pass [0.05 to 40Hz] and mains notch filters.

Pan-Tompkins algorithm is one of the most common segmentation methods used for ECG analysis \citep{PanTompkins1985,Agostinelli}. This method uses a patient-specific threshold-based approach for real-time detection of the QRS complex in ECG signals, which represents the ventricular depolarization. In this algorithm, after a pre-processing step by a band-pass filter, the signal is passed through differentiation and squaring blocks to determine and amplify the slope of QRS, followed by a moving window integration step with an adaptive set of thresholds to determine the peaks. The detection thresholds are learned at the beginning of the algorithm and are calibrated periodically to follow the variations of the ECG signal.

\begin{figure}[!ht]
 
    \centering{
\begin{tikzpicture}[>=latex']
\clip (-1,-4) rectangle (14, 9);
 \node[draw,circle,fill=blue!20, minimum size=1cm] at (1, 5.5) (O1) {$O_1$};
 \node[draw,circle,fill=blue!50, minimum size=1cm] at (2.5, 4.4) (Q) {$Q$};
 \node[draw,circle,fill=blue!50, minimum size=1cm] at (4, 7) (R) {$R$};
 \node[draw,circle,fill=blue!50, minimum size=1cm] at (4.5, 0.5) (S) {$S$};
 \node[draw,circle,fill=blue!20, minimum size=1cm] at (6, 3) (O2) {$O_2$};
  \node[draw,circle,fill=blue!20, minimum size=1cm] at (8, 4.5) (O3) {$O_3$};
  \node[draw,circle,fill=blue!20, minimum size=1cm] at (9.5,6) (O4) {$O_4$};
  \node[draw,circle,fill=blue!20, minimum size=1cm] at (11, 5) (O5) {$O_5$};
  \node[draw,circle,fill=blue!20, minimum size=1cm] at (12.5, 4) (O6) {$O_6$};
\draw [->, line width = 1pt] (O1) to 
			node[anchor=center, sloped, anchor=center, below=0]{$\beta$} 
		    (Q) ;
\draw [->, line width = 1pt] (O1) to 
			node[rotate=90, above =0mm, shift={(10mm,-4mm)}]{$\Delta \mu_t \le 0$} 
		    (Q) ;
\draw [->, line width = 1pt] (Q) to 
			node[sloped, anchor=center, above=0]{$\Delta \mu_t \ge 2$} 
		    (R) ;
 \draw [->, line width = 1pt] (R) to 
			node[sloped, anchor=center, above=0]{$\Delta \mu_t \le 5$} 
		    (S) ;	  
\draw [->, line width = 1pt] (S) to 
			node[sloped, anchor=center, above=0]{$\Delta \mu_t \ge 2$} 
		    (O2) ;	  
\draw [->, line width = 1pt] (O2) to 
			node[sloped, anchor=center, above=0]{$\Delta \mu_t \ge 1$} 
		    (O3) ;
\draw [->, line width = 1pt] (O3) to 
			node[rotate=90, above =0mm, shift={(8mm,0mm)}]{$\Delta \mu_t \ge 0$} 
		    (O4) ;
\draw [->, line width = 1pt] (O4) to 
			node[rotate=90, above =0mm, shift={(10mm,-4mm)}]{$\Delta \mu_t \le 0$} 
		    (O5) ;
\draw [->, line width = 1pt] (O5) to 
			node[rotate=90, above =0mm, shift={(10mm,-4mm)}]{$\Delta \mu_t \le 0$} 
		    (O6) ;
		    (O7) ;
		    
\draw [->, bend right = 90 , line width = 1pt] (O6) to 
			node[sloped, anchor=center, below=0]{$\Delta \mu_t \ge 0$} 
		    (O1) ;

		 \draw [->, dashed, loop below,line width = 1pt] (O1) to node[midway, above=1mm]{}  (O1) ;
		\draw [->, dashed, loop below,line width = 1pt] (Q) to node[midway, below=1mm]{} (Q) ;
		\draw [->, dashed, loop right,line width = 1pt] (R) to node[midway, right=1mm]{} (R) ;
		\draw [->, dashed, loop left,line width = 1pt] (S) to node[midway, left=1mm]{} (S) ;
        \draw [->, dashed, loop below,line width = 1pt] (O2) to node[midway, below=1mm]{} (O2) ;
        \draw [->, dashed, loop below,line width = 1pt] (O3) to node[midway, below=1mm]{} (O3) ;
        \draw [->, dashed, loop below,line width = 1pt] (O4) to node[midway, below=1mm]{} (O4) ;
        \draw [->, dashed, loop below,line width = 1pt] (O5) to node[midway, below=1mm]{} (O5) ;
        \draw [->, dashed, loop below,line width = 1pt] (O6) to node[midway, below=1mm]{} (O6) ;
\end{tikzpicture}
}
\vskip -4.3cm
\caption{Graph structure of proposed nine-state constrained change-point model. The graph is cyclic: the last node $O_1$ is the first node $O_1$. Only one transition (from $O_1$ to $Q$) to enter the QRS complex is penalized by a positive penalty $\beta=8\times 10^3$. We used the notation $\Delta \mu_t = \mu_{t+1}-\mu_t$. Transitions from state $Q$ to state $O_3$ are constrained with a minimal gap size of $2$, $5$, $2$ and $1$. Due to lack of space, we removed the indicator function $I$ on this graph. Dashed arrows correspond to $I_{\mu_t=\mu_{t+1}}$ transitions. The vertical position of the states gives information on the direction of the constrained changes.}
\label{fig:graphecg}
\end{figure}
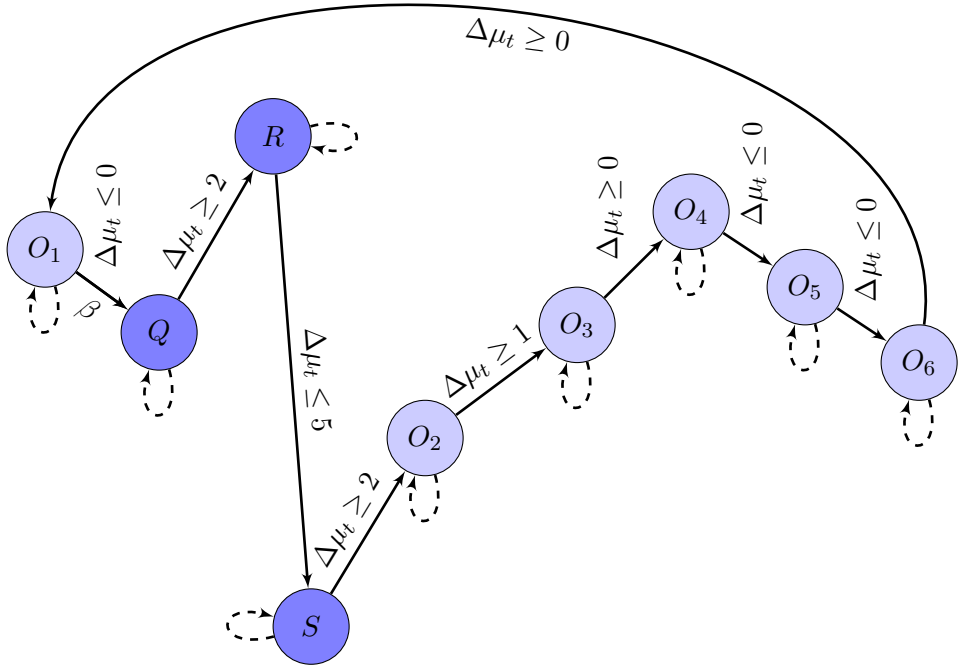

\begin{figure}[!ht]
\centering
    \includegraphics[width=\textwidth]{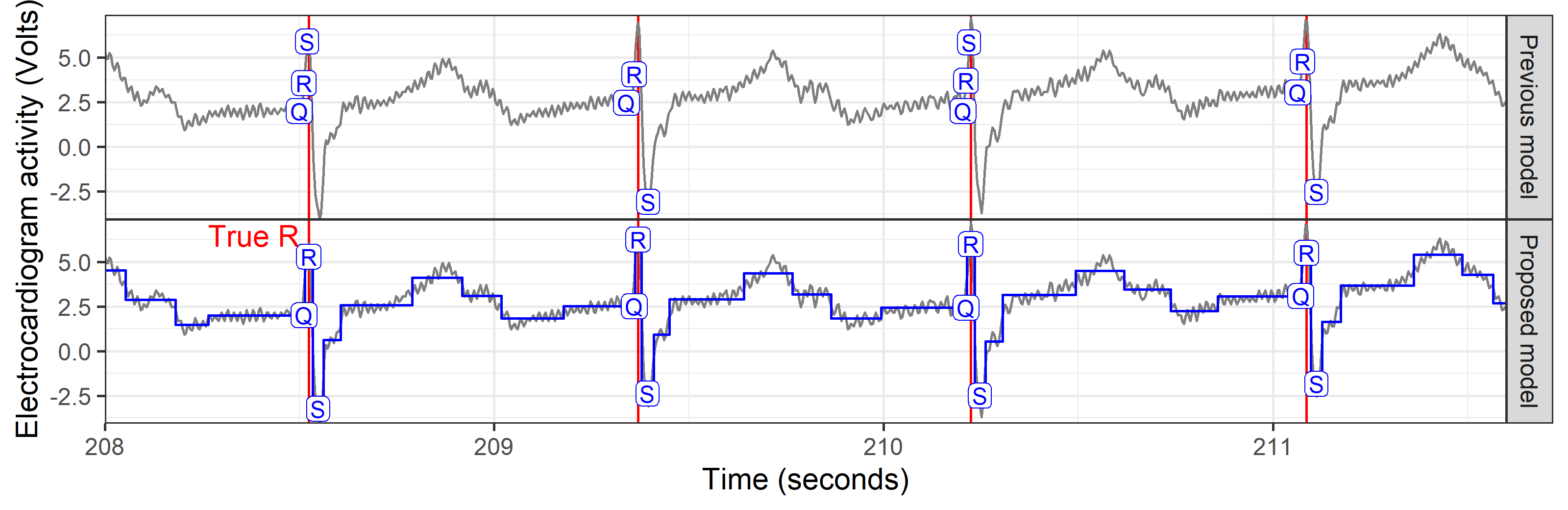}
    \vskip -0.5cm
\caption{In these electrocardiogram data, it is important for models (\textcolor{blue}{blue}) to accurately detect the QRS complex (Q is before the peak, R is the peak marked in red, S is the local minimum after the peak, other states o1--o6). \textbf{Top:} Previous model of \citet{PanTompkins1985} mistakenly predicts S at the peak. 
\textbf{Bottom:} proposed constrained change-point model accurately predicts R at each peak.}
    \label{fig:ecg}
\end{figure}

Figure~\ref{fig:ecg} (top) shows four seconds of ECG data for which we predicted the QRS complex using the well-known Pan-Tompkins method. The peak of each heartbeat should be predicted as R, but the algorithm incorrectly predicts S in two cases. In contrast the peak is correctly classified as R (bottom) using our proposed model with nine states (see Figure~\ref{fig:graphecg}), which were determined using prior knowledge about the expected sequence of changes. In this model, the QRS complex is modeled by an up-spike followed by a down spike with the maximum amplitude difference related to adjacent spikes. The graph model considers a vertex for each main waveform(i.e., P, Q, R, S, T) as well as three baselines, which are intermediate states \citep{FOTOOHINASAB2021104208}.

\section{Isotonic regression using constraint graph with robust loss}\label{sec:iso}

Our package can be used with robust loss functions which have been shown to be useful in the presence of outliers \citep{fearnhead2018changepoint} and in particular in the context of isotonic regression \citep{bach2018efficient}.
Here we illustrate this on simulations inspired by those of \cite{bach2018efficient} (see Figure~\ref{fig:iso-simu-example} with corrupted data).
We compare our package using the isotonic model described in Figure~\ref{fig:iso_col} with several implementations of the PAVA
algorithm \citep{best1990active,de2010isotone}.

Relative to the very fast $O(n)$ PAVA, our dynamic programming algorithm is slower. However, PAVA only works for the square loss and the non-penalized model (maximum number of changes). In contrast, \fct{gfpop} can handle non-convex losses (such as the biweight loss) and can include a positive penalty in order to reduce the number of changes.


\subsection{Parametrization}

\paragraph{\pkg{gfpop}.}
In all simulations we used \pkg{gfpop} with the graph of Figure~\ref{fig:iso_col} and a quadratic (L2) or a biweight loss (bw). We considered two different values for the penalty $\beta$: $0$ and $2\sigma^2\log(n)$, with $\sigma^2$ the true variance. Thus, we have 4 different algorithms of the \fct{gfpop} function: \code{gfpop1} ($\beta = 0$, $K = 0$),  \code{gfpop2} ($\beta = 2\sigma^2 \log(n)$, $K = 0$),  \code{gfpop3} ($\beta = 0$, $K = 3\sigma$) and  \code{gfpop4} ($\beta = 2\sigma^2 \log(n)$, $K = 3\sigma$).

\paragraph{Competitors.}
We compared the output of \pkg{gfpop} with those of 2 isotonic regression package functions:
\begin{itemize}
    \item \fct{isoreg} function of the \pkg{stats} package which is based on the very fast Pool adjacent violators algorithm for the $\ell_2$ loss  \citep{best1990active};
    \item \code{reg}\underline{\hspace{0.2cm}}\fct{1d} function developed in package \pkg{UniIsoRegression} which solves the isotonic regression problem for the $\ell_2$ and $\ell_1$ losses \citep{stout2008unimodal}.
\end{itemize}

We also include a simple linear regression approach (\fct{lm} function of the \pkg{stats} package) as a reference. In total we have 4 competitors (\fct{lm}, \fct{isoreg}, \code{reg}\underline{\hspace{0.2cm}}\fct{1d} with the $\ell_2$ and $\ell_1$ losses). 


\subsection{Simulated data}
 We focused on two types of increasing signals:
 \begin{description}
 \item[linear:] as in \cite{bach2018efficient} we consider linearly increasing time series with a signal 
 $$ s_i = \alpha (i-\frac{n}{2})\,\quad i = 1,...,n\,;$$
 \item[step-wise:] as our package is devoted to change-point inference we also consider a step-wise increasing series (with $10$ steps) with a signal 
 $$s_i = \lfloor \frac{10(i-1)}{n} \rfloor - \frac{n}{2}\,,\quad i = 1,...,n\,.$$ 
 \end{description}

 We consider three ways to corrupt the data.

 \begin{description}
 \item[Gaussian noise:] here we simply add a Gaussian noise, with a variance $\sigma^2$ to the signal (i.e., $Y_i=s_i+\varepsilon_i$).
 \item[Student noise:] we also considered a Student noise with a degree of freedom equal to $3$.
 \item[Corrupted noise:] in the most difficult scenario, suggested by \cite{bach2018efficient}, we randomly select a proportion $p$ of data-points and multiply them by $-1$ and then add a Gaussian noise, i.e., $Y_i= X_i s_i + \varepsilon_i$, where $X_i \sim \mathcal{B}(p)$ is a Bernoulli trial with probability $p$ to get $-1$ and probability $1-p$ to get $1$. We fix $p = 0.3$ for all simulations.
 \end{description}

\begin{figure}[ht!]
\centering
\includegraphics[width=0.7\columnwidth]{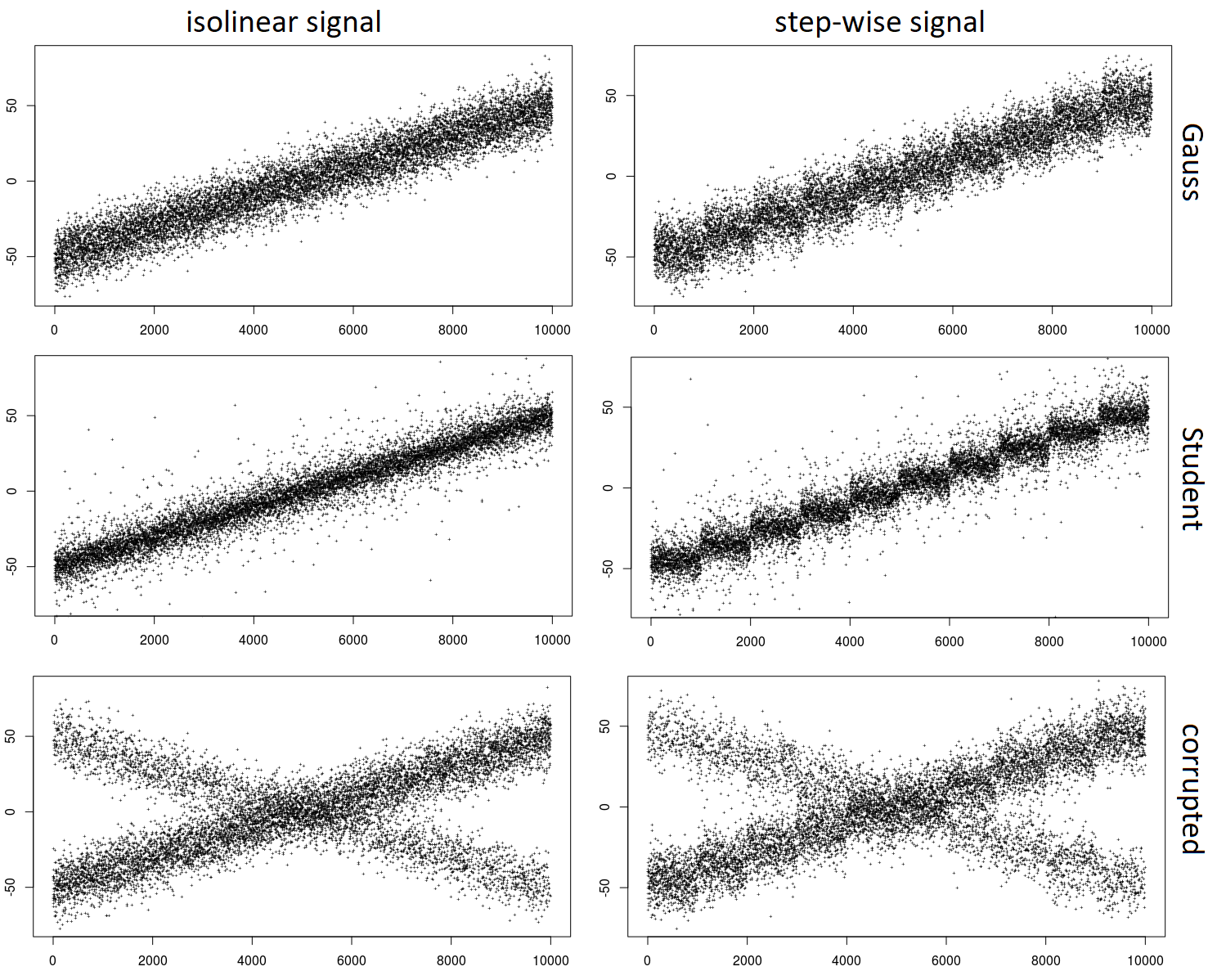}
\caption{\label{simu6cases}For the two types of signal, we show simulated data with $n = 10^4$ data-points and $\sigma = 10$ for the three different noises (Gauss, Student, corrupted).}\label{fig:iso-simu-example}
\end{figure}

In total we have 6 scenarios (2 signals and 3 ways to corrupt the data).
In Figure~\ref{fig:iso-simu-example} we illustrate those 6 scenarios with $n = 10^4$ and $\sigma = 10$, which is an example of time series used in following simulations.

\paragraph{Criteria.} To assess the quality of the results, we compute the Mean-Squared Error (MSE) as well as the ability to recover the true number of changes when there are changes in the data in the step-wise scenario.


\subsection{A simple illustration}

\begin{figure}[ht!]
\centering
\includegraphics[width=0.6\columnwidth]{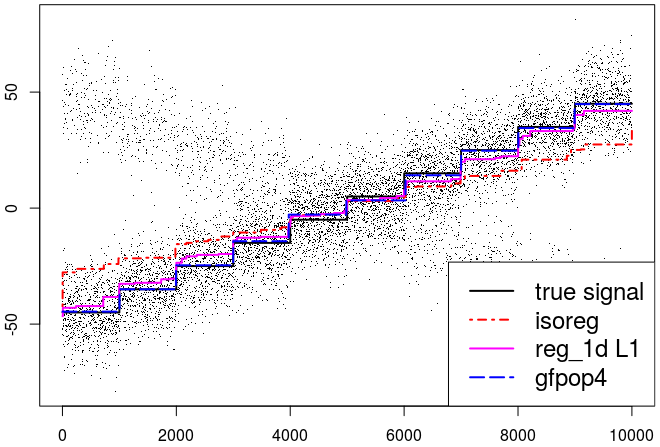}
\caption{\label{isotonic}Isotonic regression with $30\%$ of corrupted data in step-wise scenario with $10$ steps. We have $10^4$ data-points and $\sigma = 10$. \fct{gfpop4} is close to the signal and with a number of segments equal to $10$.}
\end{figure}

We illustrate our results on a step-wise increasing signal with corrupted data. In Figure~\ref{isotonic}, we represent the data and the results of various approaches. We see that using a biweight loss our package in blue is closer to the true signal in black than other approaches.

In Appendix we considered Monte-Carlo simulations to confirm this result: see Section \ref{subsec:isolin} for the linear increasing scenario and Section \ref{subsec:isostep} for the step-wise increasing scenario. As expected, recovering the true number of changes with corrupted data with also a small MSE is a challenging task for all methods excepted for \fct{gfpop4} using a robust loss, a positive penalty and the isotonic constraint graph. The \proglang{R} code of these simulations can be found on the Github page \url{https://github.com/vrunge/gfpop/tree/master/simulations}.


\section{Conclusion}

In this paper we described the \pkg{gfpop} package, which provides a generalized version of an algorithm recently proposed by \cite{hocking2020constrained} for penalized maximum likelihood inference of constrained multiple change-point models. The \pkg{gfpop} package implements the algorithm in a generic manner in \proglang{R/C++} and allows the user to specify the constraint graph in \proglang{R} code. We explained how these constrained multiple change-point models can also be seen as constrained continuous state space HMM.
\pkg{gfpop} allows one to encode modeling assumptions on the type of changes using a graph of states and constraints. We illustrated the use of \pkg{gfpop} on isotonic simulations and several applications in biology. 

For a number of graphs the algorithm runs in a matter of seconds or minutes for $10^5$ data-points. While \pkg{gfpop} can be used to fit simple change-point models, such as the standard change in mean in Gaussian data, it is slower than \pkg{fpop} which implements functional programming specifically for that model: For example, for $10^5$ points with no change \pkg{fpop} runs in $0.031$ seconds and \pkg{gfpop} in $0.35$ seconds. This is because the \pkg{gfpop} is coded in a more generic manner, as it handles constraints and various losses. As we illustrated with numerous examples, the advantage of \pkg{gfpop} is that it allows one to include constraints and/or unconventional losses, and thus fit a range of change-point models that cannot be fit by other generic software.

\paragraph{Future Work.} For future work we are interested to explore generalizations which allow time-dependent constraints. 
As mentioned in Section~\ref{subsec:collapsed} our implementation only allows inference in models that can be represented by a collapsed graph with transitions that are valid for all time points.
We are interested in exploring new frameworks for defining which transitions and/or states are feasible at which time points, in order to efficiently support inference in models such as Labeled Optimal Partitioning \citep{Hocking2020-LOPART}. There are a number of other extensions of \pkg{gfpop} that are possible, including allowing local fluctuations in the parameter between change-points and modeling auto-correlated noise -- these can be both be incorporated using ideas from \cite{romano2020detecting}. Another extension would be to consider the detection of change-points in trees as proposed in chapter three of the Ph.D. thesis of \citet{thep2019}. Furthermore, the underlying \code{gfpop} algorithm is sequential and thus can be adapted to allow for online change-point detection.


\bibliographystyle{unsrtnat}
\bibliography{jss4384}
\newpage


\appendix

\section{Some other graphs}\label{app-sec:graphs}

Here are three graphs for models discussed in the main part of the
paper.

\begin{figure}[!ht]
\centering{
\begin{tikzpicture}[>=latex']
 \node[draw,circle, minimum size=1cm] at (0, 0) (step_0) {$\#$};
 \node[draw,circle,, minimum size=1cm] at (8, -3) (step_{n+1}) {$\emptyset$};
 \node[draw,circle,fill=blue!20, minimum size=1cm] at (12, 0) (state_Coll) {Coll};
 \node[draw,circle,fill=blue!20, minimum size=1cm] at (4, 0) (state_Norm) {$\mu_0$};
 
 \draw [->, dotted, line width = 1pt] (step_0) to (state_Norm) ;
 \draw [->, dashed, loop above, line width = 1pt] (state_Norm) to 
			node[midway, above=0]{$I_{\mu_t = \mu_{t+1}}$, $bw$} 
		    (state_Norm) ;
 \draw [->, bend left, line width = 1pt] (state_Norm) to 
			node[midway, above=0] {$I_{\mu_t \neq \mu_{t+1}}$, $\ell_2$, $\beta$} 
		    (state_Coll) ;
 \draw [->, dotted, line width = 1pt] (state_Norm) to (step_{n+1}) ;
 
 \draw [->, dashed, bend left, line width = 1pt] (state_Coll) to 
			node[midway, above=0]{$I_{\mu_t \neq \mu_{t+1}}$, $bw$} 
		    (state_Norm) ;
\draw [->, dashed, loop above, line width = 1pt] (state_Coll) to 
			node[midway, above=0]{$I_{\mu_t = \mu_{t+1}}$, $\ell_2$} 
		    (state_Coll) ;
\draw [->, dotted, line width = 1pt] (state_Coll) to (step_{n+1}) ;
 
\end{tikzpicture}
}
\caption{Graph for the model proposed in \cite{fisch2018linear} 
We have $\mathcal{S}=\{\mu_0, Coll\}$. Note that the value of $\mu_0$ is given and that the loss function is either the $\ell_2$ or the biweight $bw$. The penalty is omitted when equal to zero.
}
\label{fig:collective}
\end{figure}

\begin{figure}[!ht]
\centering{
\begin{tikzpicture}[>=latex']
 \node[draw,circle, minimum size=1cm] at (0, 0) (step_0) {$\#$};
 \node[draw,circle, minimum size=1cm] at (12, 0) (step_{n+1}) {$\emptyset$};
 \node[draw,circle,fill=blue!20, minimum size=1cm] at (9, 0) (state_Seg) {Seg};
 \node[draw,circle,fill=blue!20, minimum size=1cm] at (3, 0) (state_Wait1) {Wait$_1$};
 \node[draw,circle,fill=blue!20, minimum size=1cm] at (6, -2) (state_Wait2) {Wait$_2$};
 \draw [->, dotted,line width = 1pt] (step_0) to (state_Wait) ;
 \draw [->, dashed, loop above,line width = 1pt] (state_Seg) to 
			node[midway, above=0]{$I_{\mu_t = \mu_{t+1}}$} 
		    (state_Seg) ;
 \draw [->,line width = 1pt] (state_Seg) to 
			node[midway, above=0] {$I_{\mu_t \neq \mu_{t+1}}$, $\beta$} 
		    (state_Wait1) ;
 \draw [->, dotted,line width = 1pt] (state_Seg) to (step_{n+1}) ;
 
 \draw [->, dashed,line width = 1pt] (state_Wait1) to 
			node[midway, below=0, sloped]{$I_{\mu_t = \mu_{t+1}}$} 
		    (state_Wait2) ;
\draw [->, dashed,line width = 1pt] (state_Wait2) to 
			node[midway, below=0, sloped]{$I_{\mu_t = \mu_{t+1}}$} 
		    (state_Seg) ;
\end{tikzpicture}
}
\caption{Graph for the at least 3 data-point per segment model. We have $\mathcal{S}=\{Wait_1, Wait_2, Seg\}$, the loss function is always the $\ell_2$. The penalty is omitted when equal to zero.}
\label{fig:atleast3}
\end{figure}
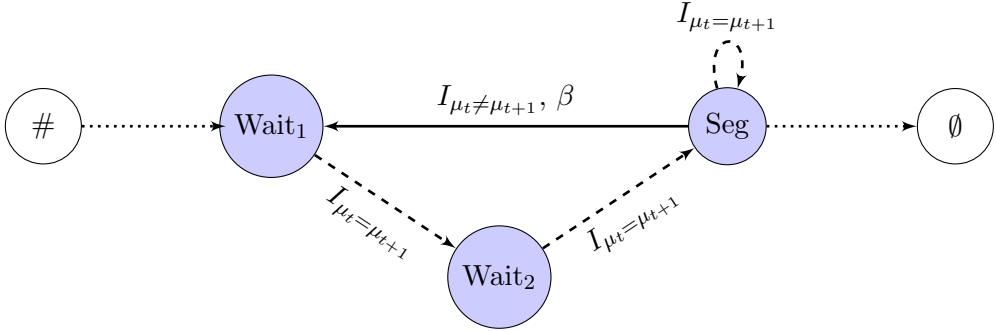

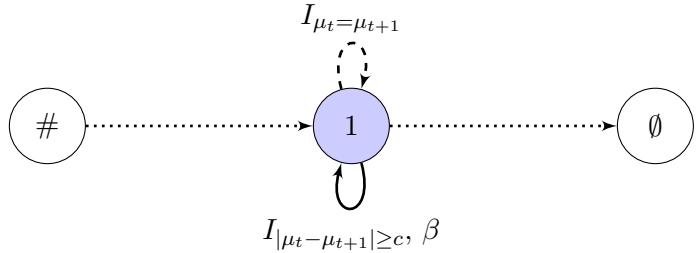
\begin{figure}[!ht]
\centering{
\begin{tikzpicture}[>=latex']
 \node[draw,circle, minimum size=1cm] at (0, 0) (step_0) {$\#$};
 \node[draw,circle, minimum size=1cm] at (8, 0) (step_{n+1}) {$\emptyset$};
 \node[draw,circle,fill=blue!20, minimum size=1cm] at (4, 0) (state_1) {$1$};
 \draw [->, dotted, line width = 1pt] (step_0) to (state_1) ;
 \draw [->, dashed, loop above, line width = 1pt] (state_1) to 
			node[midway, above=0]{$I_{\mu_t = \mu_{t+1}}$} 
		    (state_1) ;
 \draw [->, loop below=0cm, line width = 1pt] (state_1) to 
			node[midway, below=0]{$I_{|\mu_t - \mu_{t+1}| \geq c }$, $\beta$} 
		    (state_1) ;
 \draw [->, dotted, line width = 1pt] (state_1) to (step_{n+1}) ;

\end{tikzpicture}
}
\caption{Graph for relevant change-point model. We have $\mathcal{S}=\{1\}$, the loss function is always the $\ell_2$. The penalty is omitted when equal to zero.
}
\label{fig:relevant}
\end{figure}

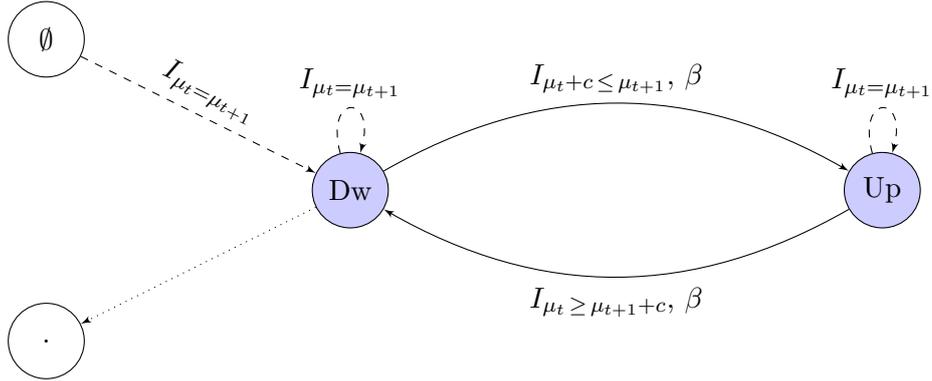
\begin{figure}[!ht]
\centering{
\begin{tikzpicture}[>=latex']
 \node[draw,circle, minimum size=1cm] at (0, 0) (step_0) {$\emptyset$};
 \node[draw,circle, minimum size=1cm] at (0, -4) (step_{n+1}) {$.$};
 \node[draw,circle,fill=blue!20, minimum size=1cm] at (11, -2) (state_Up) {Up};
 \node[draw,circle,fill=blue!20, minimum size=1cm] at (4, -2) (state_Dw) {Dw};
 
 \draw [->, dashed] (step_0) to 
			node[midway, sloped, above=0]{$I_{\mu_t = \mu_{t+1}}$} 
		    (state_Dw) ;
 \draw [->, dashed, loop above] (state_Up) to 
			node[midway, above=0]{$I_{\mu_t = \mu_{t+1}}$} 
		    (state_Up) ;
 \draw [->, bend left] (state_Up) to 
			node[midway, below=0] {$I_{\mu_t \,\geq \,  \mu_{t+1} + c}$, $\beta$} 
		    (state_Dw) ;

 \draw [->, dashed, loop above] (state_Dw) to 
			node[midway, above=0]{$I_{\mu_t = \mu_{t+1}}$} 
		    (state_Dw) ;
\draw [->, bend left] (state_Dw) to 
			node[midway, above=0]{$I_{ \mu_t + c \,\leq \, \mu_{t+1}}$, $\beta$} 
		    (state_Up) ;
\draw [->, dotted] (state_Dw) to (step_{n+1}) ;
 
\end{tikzpicture}
}
\caption{Graph for the Up-Down relevant model.We have $\mathcal{S}=\{Up, Dw\}$, the loss function is always the $\ell_2$. The penalty is omitted when equal to zero.
}
\label{fig:updownRel}
\end{figure}

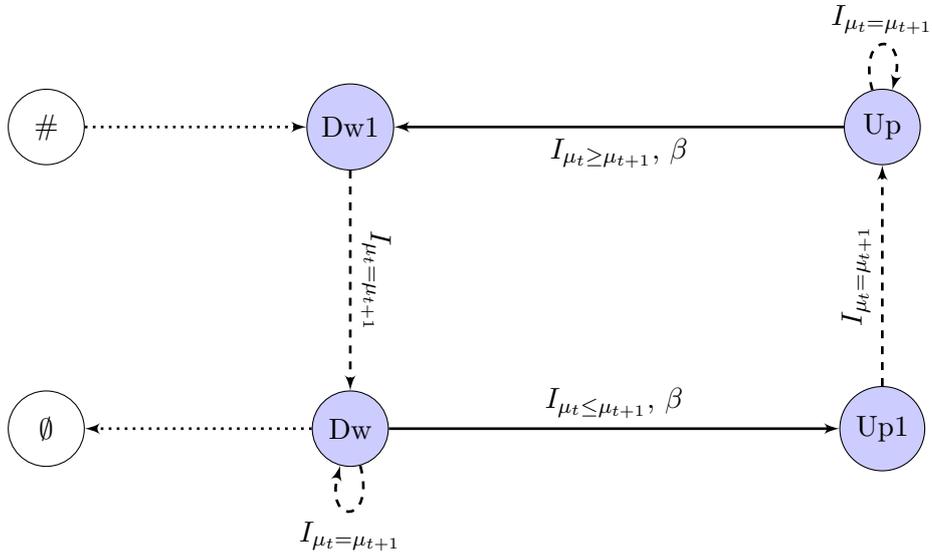
\begin{figure}[!ht]
\centering{
\begin{tikzpicture}[>=latex']
 \node[draw,circle, minimum size=1cm] at (0, 0) (step_0) {$\#$};
 \node[draw,circle, minimum size=1cm] at (0, -4) (step_{n+1}) {$\emptyset$};
 \node[draw,circle,fill=blue!20, minimum size=1cm] at (11, -4) (state_Up1) {Up1};
 \node[draw,circle,fill=blue!20, minimum size=1cm] at (4, 0) (state_Dw1) {Dw1};
 \node[draw,circle,fill=blue!20, minimum size=1cm] at (11, 0) (state_Up) {Up};
 \node[draw,circle,fill=blue!20, minimum size=1cm] at (4, -4) (state_Dw) {Dw};

 \draw [->, dotted, line width = 1pt] (step_0) to (state_Dw1) ;
 \draw [->, dashed, line width = 1pt] (state_Up1) to 
			node[midway, above=0, sloped]{$I_{\mu_t = \mu_{t+1}}$} 
		    (state_Up) ;

 \draw [->, dashed, loop above, line width = 1pt] (state_Up) to 
			node[midway, above=0]{$I_{\mu_t = \mu_{t+1}}$} 
		    (state_Up) ;
 \draw [->, line width = 1pt] (state_Up) to 
			node[midway, below=0] {$I_{\mu_t \geq \mu_{t+1}}$, $\beta$} 
		    (state_Dw1) ;

 \draw [->, dashed, line width = 1pt] (state_Dw1) to 
			node[midway, above=0, sloped]{$I_{\mu_t = \mu_{t+1}}$} 
		    (state_Dw) ;
 \draw [->, dashed, loop below, line width = 1pt] (state_Dw) to 
			node[midway, below=0]{$I_{\mu_t = \mu_{t+1}}$} 
		    (state_Dw) ;
\draw [->, line width = 1pt] (state_Dw) to 
			node[midway, above=0]{$I_{\mu_t \leq \mu_{t+1}}$, $\beta$} 
		    (state_Up1) ;
\draw [->, dotted, line width = 1pt] (state_Dw) to (step_{n+1}) ;
 
\end{tikzpicture}
}
\caption{Graph for the Up-Down model with segments of size at least 2. We have $\mathcal{S}=\{\text{Up1}, \text{Up}, \text{Dw1}, \text{Dw} \}$. The loss function is always the $\ell_2$ or the Poisson. The penalty is omitted when equal to zero.
}
\label{fig:updown2}
\end{figure}

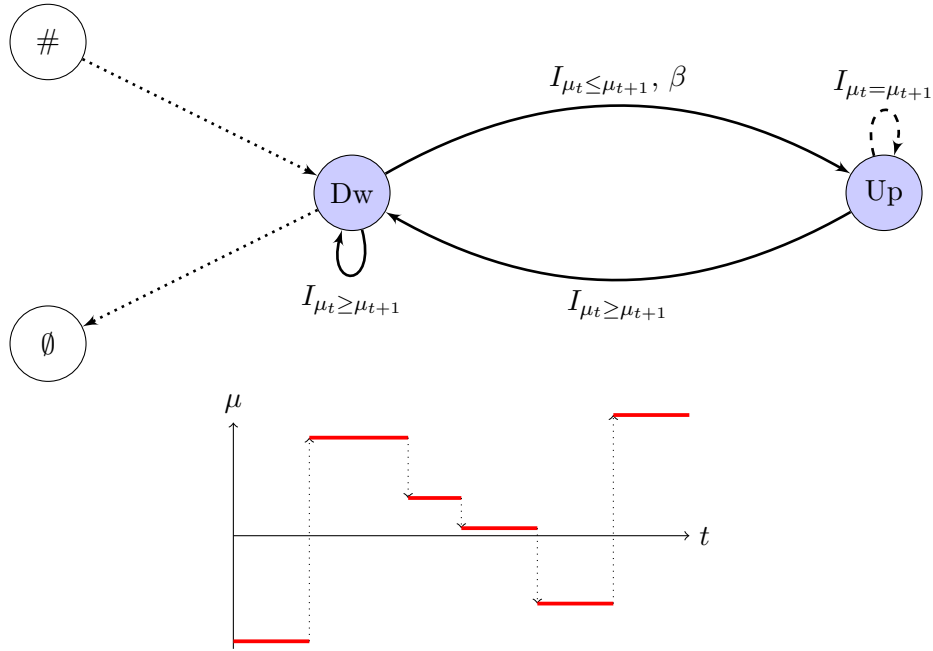
\begin{figure}[!ht]
\centering{
\begin{tikzpicture}[>=latex']
 \node[draw,circle, minimum size=1cm] at (0, 0) (step_0) {$\#$};
 \node[draw,circle, minimum size=1cm] at (0, -4) (step_{n+1}) {$\emptyset$};
 \node[draw,circle,fill=blue!20, minimum size=1cm] at (11, -2) (state_Up) {Up};
 \node[draw,circle,fill=blue!20, minimum size=1cm] at (4, -2) (state_Dw) {Dw};
\draw [->, dotted, line width = 1pt] (step_0) to
		    (state_Dw) ;

 \draw [->, bend left,line width = 1pt] (state_Up) to 
			node[midway, below=0] {$I_{\mu_t \geq \mu_{t+1}}$} 
		    (state_Dw) ;
\draw [->, loop below,line width = 1pt] (state_Dw) to 
			node[midway, below=0]{$I_{\mu_t \geq \mu_{t+1}}$} 
		    (state_Dw) ;
\draw [->, bend left,line width = 1pt] (state_Dw) to 
			node[midway, above=0]{$I_{\mu_t \leq \mu_{t+1}}$, $\beta$} 
		    (state_Up) ;
\draw [->, dotted,line width = 1pt] (state_Dw) to (step_{n+1}) ;
 
\end{tikzpicture}
}
\caption{(Top) Graph for the up-down* change-point model. We have $\mathcal{S}=\{Dw, Up\}$, the loss function is always the $\ell_2$. The penalty is omitted when equal to zero. (Bottom) In red, a piecewise constant function validating the graph of constraints.}
\label{fig:updownstar_col}
\end{figure}

\begin{figure}[!ht]
\centering{
\begin{tikzpicture}[>=latex']
 \node[draw,circle, minimum size=1cm] at (0, 0) (step_0) {$\#$};
 \node[draw,circle, minimum size=1cm] at (0, -4) (step_{n+1}) {$\emptyset$};
 \node[draw,circle,fill=blue!20, minimum size=1cm] at (11, -2) (state_Up) {Up};
 \node[draw,circle,fill=blue!20, minimum size=1cm] at (4, -2) (state_Dw) {Dw};
\draw [->, dotted, line width = 1pt] (step_0) to
		    (state_Dw) ;
 \draw [->, bend left,line width = 1pt] (state_Up) to 
			node[midway, below=0] {$I_{\mu_t \geq \mu_{t+1}}$, $\beta$} 
		    (state_Dw) ;

\draw [->, loop below,line width = 1pt] (state_Up) to 
			node[midway, below=0]{$I_{\mu_t \leq \mu_{t+1}}$} 
		    (state_Up) ;
		    
\draw [->, loop below,line width = 1pt] (state_Dw) to 
			node[midway, below=0]{$I_{\mu_t \geq \mu_{t+1}}$} 
		    (state_Dw) ;
\draw [->, bend left,line width = 1pt] (state_Dw) to 
			node[midway, above=0]{$I_{\mu_t \leq \mu_{t+1}}$, $\beta$} 
		    (state_Up) ;
\draw [->, dotted,line width = 1pt] (state_Dw) to (step_{n+1}) ;
 
\end{tikzpicture}
}
\caption{Graph for the up*-down* change-point model. We have $\mathcal{S}=\{Dw, Up\}$, the loss function is always the $\ell_2$. The penalty is omitted when equal to zero.}
\label{fig:upstardownstar_col}
\end{figure}
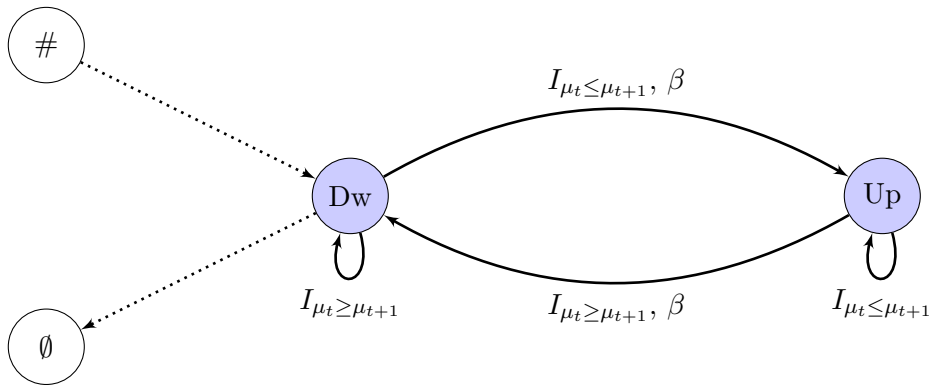

Below we provide a few other constraint models and their graphs. 
\begin{itemize}
\item (Up - Down Relevant) It might make sense to consider sufficiently large changes. This is a simple modification of the Up - Down model (see Figure~\ref{fig:updown_col}). The $Dw$ to $Up$ constraint $I_{\mu_t \leq \mu_{t+1}}$ can be replaced by $I_{c + \mu_t \leq  \mu_{t+1}}$ for $c > 0$ or $I_{a \mu_t \leq \mu_{t+1}}$ for $a > 1$ if $\mu_t$ are positive. The graph is shown in Figure~\ref{fig:updownRel}.

\item (Up - Down with at least two data-points) If one wants to detect peaks and is certain that segments are at least of length 2 it suffices to add two waiting states in the Up - Down graph. The graph of this model is given Figure~\ref{fig:updown2}.

\item (Up - Isotonic Down) In the pulse detection example (Up - Exponentially Down model in Figure~\ref{fig:updownexp_col}) if one is not sure of the exponential decrease it could make sense to consider an isotonic decrease. For this it suffices to consider two states $\mathcal{S} = \{Up, Dw\}$. Compared to the 
Up - Down model, described earlier, we add an additional transition from $Dw$ to $Dw$ with the constraint $I_{\mu_t \geq \mu_{t+1}}$. The graph of this model is given in Figure~\ref{fig:updownstar_col}.

\item (Isotonic Up - Isotonic Down) In the previous model
one considers a sharp transition up. It might make sense
to consider an isotonic increase. For this it suffices
to add an edge from $Up$ to $Up$ in the previous model. Only transitions from $Up$ to $Dw$ and $Dw$ to $Up$ are penalized.
The graph of this model is given in Figure~\ref{fig:upstardownstar_col}.

\end{itemize}


\newpage
~\\
\newpage

\section{Update-rule proof}\label{app-sec:proof-update}

We recall here the update-rule \eqref{eq:update_rule} given at the end of Section \ref{sec:opandupdate}.

\begin{equation*}
 Q_{n+1}^{s'} (\theta) = \min_{s | \exists \,edge \,(s, s')} \left\{ O_n^{s,s'}(\theta) + \gamma_{(s, s')}(y_{n+1},\theta) + \beta_{(s, s')}  \right\}\,.
\end{equation*}

We name the path and vector realizing the best cost $Q_{n+1}^{s'} (\theta)$, defined in Equation~\ref{eq:func_cost_opt}, $p^*$ and $\mu^*$.
We call $s^*$ the corresponding vector of states. We have $s^*_{n+1}=s'$, $\mu^*_{n+1}=\theta$ and 
$$Q_{n+1}^{s'} (\theta) =  \sum_{t=1}^{n+1} \left(\gamma_{e^*_t}(y_t, \mu^*_t) + \beta_{e^*_t}\right)  = \sum_{t=1}^{n} \left(\gamma_{e^*_t}(y_t, \mu^*_t) + \beta_{e^*_t}\right) + (\gamma_{(s^*_n, s')}(y_{n+1}, \theta) + \beta_{(s^*_n, s')}) \,.$$

\paragraph{}
We will first show that
\begin{equation*}
    \sum_{t=1}^{n} \left(\gamma_{e^*_t}(y_t, \mu^*_t) + \beta_{e^*_t}\right) = Q_{n}^{s^*_n} (\mu^*_n) = O_{n}^{s^*_n, s'} (\theta). 
\end{equation*}

(Proof) Restricting the path $p^*$ and the vector $\mu^*$ to their first $n$ elements, by definition of $Q_{n}^{s^*_n} (\mu^*_n)$ we have 
$\sum_{t=1}^{n} \left(\gamma_{e^*_t}(y_t, \mu^*_t) + \beta_{e^*_t}\right) \geq Q_{n}^{s^*_n} (\mu^*_n).$
Also, given that a move from parameter $\mu^*_n$ to  $\theta$ is a valid transition from state $s^*_n$ to $s'$ and by the definition of $O_{n}^{s^*_n, s'}(\theta)$,  we have $Q_{n}^{s^*_n} (\mu^*_n) \geq O_{n}^{s^*_n, s'}(\theta).$

We will now proceed by contradiction.
Let us assume that $ \sum_{t=1}^{n} \left(\gamma_{e^*_t}(y_t, \mu^*_t) + \beta_{e^*_t}\right) > O_{n}^{s^*_n, s'} (\theta)$.
We name the path and vector realizing the
$O_{n}^{s^*_n, s'} (\theta)$ 
 $p^+$ and $\mu^+.$
Extending this path and vector to $n+1$ with 
$s^+_{n+1}=s'$ and $\mu^+_{n+1}=\theta$ we get a better cost than $p^*$ for  $Q_{n+1}^{s'} (\theta)$ which is a contradiction.

So we have
\begin{equation*}
 Q_{n+1}^{s'} (\theta) =  O_n^{s^*_n,s'}(\theta) + \gamma_{(s, s')}(y_{n+1},\theta) + \beta_{(s^*_n, s')}, 
\end{equation*}
and considering all possible states at time $n$ we get the update-rule.

\newpage


\section{Backtracking}
\label{app-sec:backtracking}

After running the Viterbi-like algorithm with update-rule \ref{eq:update_rule}, we need a backward procedure called backtracking to return the optimal change-point vector. First, we recover using Algorithm \ref{algo1} the optimal vector of states $\hat{s} \in \{1,...,S\}^n$ and vector of means $\hat{\mu} \in \mathbb{R}^n$. We then find the best change-point vector $\hat{\tau} \subset \{1,...,n\}$ with Algorithm \ref{algo2}. The basic idea of Algorithm \ref{algo1} is that if we knew $\hat{s}_{t+1}$ and $\hat{\mu}_{t+1}$ we could recover first $\hat{s}_t$ and then $\hat{\mu}_{t}$ taking the argmin of the update-rule (see lines 8 and 9 of Algorithm \ref{algo1}).

\begin{algorithm}
\caption{Backtracking $\hat{s}$ and $\hat{\mu}$}
\label{algo1}
\begin{algorithmic}[1]
\STATE {\bf procedure} \textsc{Backtrack}($(Q_1^1,...,Q_1^S),...,(Q_n^1,...,Q_n^S)$)
\STATE $\hat{\mu} \gets $ empty vector of size $n$
\STATE $\hat{s} \gets$  empty vector of size $n$
\STATE $ (\hat{s}_{n},\hat{\mu}_{n}) = \underset{(s,\mu)}{\argmin}\{ Q^s_n(\mu) \}$
\STATE\hfill $\triangleright$ {We can impose a subset of arrival states $\tilde{\mathcal{S}} \subset \{1,...,S\}$
\STATE\hfill by $(\hat{s}_{n},\hat{\mu}_{n}) \gets \{\underset{(s,\mu)}{\argmin}\{ Q^{s}_n(\mu)\}\,,\, s \in \tilde{\mathcal{S}}\}$}
\FOR{$t=n-1$ \TO $t=1$}
\STATE $\hat{s}_{t} = \underset{s | \exists \,edge \,(s, \hat{s}_{t+1})}{\argmin} \left\{ O_{t}^{s,\hat{s}_{t+1}}(\hat{\mu}_{t+1}) + \gamma_{(s, \hat{s}_{t+1})}(y_{t+1},\hat{\mu}_{t+1}) + \beta_{(s, \hat{s}_{t+1})}  \right\}$
\STATE$ \hat{\mu}_{t}= \underset{\mu | I_{(\hat{s}_{t},\hat{s}_{t+1})}(\mu, \hat{\mu}_{t+1}) }{\argmin} \left\{ Q^{\hat{s}_{t}}_{t}(\mu)  \right\}$ \hfill $\triangleright$ If $\hat{\mu}_{t}$ is such that the constraint is active, 
\STATE \hfill we have 'forced = TRUE' in \fct{gfpop} response
\ENDFOR
\RETURN  $(\hat{s}, \hat{\mu})$
\end{algorithmic}
\end{algorithm}

The obtained vectors $\hat{s}$ and $\hat{\mu}$ are simplified removing repetitions of consecutive identical states or values: i.e., $\hat{s}_t = \sigma_0$ and $\hat{\mu}_t = m_0 \gamma^{t_2-t}$ for $t=t_1,...,t_2$ (including the case of exponential decay with parameter $\gamma$ and $\gamma = 1$ if no decay). In that case, the index $t_2$ is an element of the change-point vector and $m_0$ its associated segment parameter. The vector of change-points can be built by a linear-in-time procedure described in Algorithm \ref{algo2}.

\begin{algorithm}
\caption{Change-point vector}
\label{algo2}
\begin{algorithmic}[1]
\STATE {\bf procedure} \textsc{Change-point}$(\hat{s}, \hat{\mu})$
\STATE $\hat{\tau} \gets NULL$
\STATE $t \gets n+1$
\WHILE{$t > 1$}
\STATE $\hat{\tau} \gets (t-1 ,\hat{\tau})$
\WHILE{$(\hat{s}_{t-1}, \gamma\hat{\mu}_{t-1}) = (\hat{s}_{t}, \hat{\mu}_{t})$} \STATE $t \gets t-1$\ENDWHILE
\ENDWHILE
\RETURN  $\hat{\tau}$
\end{algorithmic}
\end{algorithm}

Notice that $\hat{\tau}$ is the \code{changepoints} vector returned by the \fct{gfpop} function. Restricting $\hat{s}$ and $\hat{\mu}$ vectors to positions in $\hat{\tau}$, these vectors are respectively the \code{states} and the \code{parameters} vectors.

\section{Simulation results for isotonic regression}
\label{app-sec:simus}

\subsection{Linear signal}\label{subsec:isolin}

We simulate $100$ linearly increasing time series and compute the mean of the MSE for each noise structure. The results are in Table~\ref{isolineartab}. We highlight in bold the two best results in each row and also give the standard deviation (SD).

In the Gaussian case the $\ell_2$ isotonic regression and \code{gfpop1} (with $\beta=0$ and $K=\infty$) are better. For the Student and for the Corrupted scenarios the robust biweight loss with $\beta=0$ is performing better in terms of MSE. Note that it is however much slower than PAVA. Including a penalty for change-points ($\beta_0=2\sigma^2\log(n)$) deteriorates the results. This make sense as there are in fact no change-points in the data.

\begin{table}[ht!]
\begin{tabular}{|c|c|c|c|c|c|c|c|c|}
\hline
Isolinear& linear &\code{isoreg}&\code{reg}\underline{\hspace{0.2cm}}\code{1d}&\code{reg}\underline{\hspace{0.2cm}}\code{1d}&\code{gfpop1} &\code{gfpop2}&\code{gfpop3} &\code{gfpop4}\\
simulations& fit & $\ell_2$ & $\ell_1$ & $\ell_2$&$\beta = 0$ &$\beta = 0$ & $\beta = \beta_0$ & $\beta = \beta_0$\\
 & && &  & $\ell_2$ & $K = 3\sigma$& $\ell_2$ & $K = 3\sigma$ \\ \hline
Gauss&& & & & & & & \\ 
MSE& 0.0190 & \bf 0.714 &  \bf 0.712 & 1.08 & 0.826 & 0.931 & 2.60 & 3.10\\ 
(SD)& (0.020)& (0.098) & (0.098) & (0.17)& (0.15)& (0.13) & (0.27) & (0.30) \\ \hline

Student&& & & & & & & \\ 
MSE& 0.0185 & 0.683 & 0.681 & \bf 0.550 & 0.780 & \bf 0.555 & 2.56 & 2.57\\ 
(SD)& (0.19) & (0.12)& (0.12)& (0.077) & (0.17) & (0.076)& (0.24) & (0.22) \\ \hline

Corrupted&& & & & & & & \\ 
MSE& 299 & 298 & 298  & 28.7 & 294 &  \bf 4.05 & 301 & \bf 7.23 \\
(SD) & (9.1) & (8.8) & (8.8) & (2.2) & (11) & (0.63) & (8.9) & (0.89) \\ \hline

\end{tabular}
\caption{\label{isolineartab} Mean squared errors MSE $= \frac{1}{n}\sum_{i=1}^n(\hat{s}_i - s_i)^2$ for different algorithms on linear simulations with its empirical standard deviation (SD). We consider three types of noise: Gaussian, Student and corrupted.}
\end{table}


\subsection{Iso-step signal}\label{subsec:isostep}

We simulate $100$ step-wise increasing time series with $10$ segments and compute the mean of the MSE for each noise structure. The results are in Table~\ref{isosteptab}. We highlight in bold the two best results in each row and also give the standard deviation (SD).

\begin{figure}[ht!]
\centering
\includegraphics[width=0.7\columnwidth]{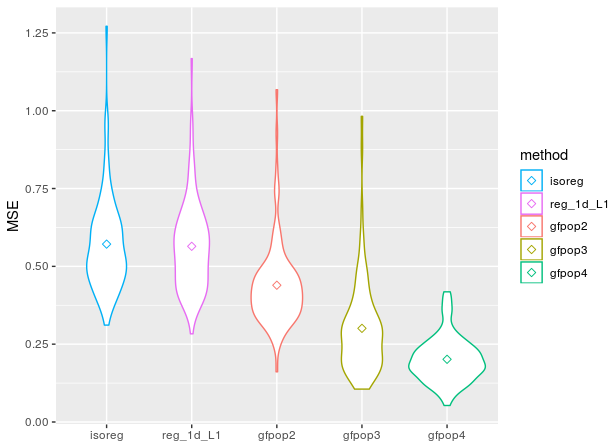}
\caption{\label{violinPlot}Violin plots of the MSE for iso-step simulations with Student noise. The shape of the distribution is very similar for the 5 best methods considered.}
\end{figure}

In Gaussian and Student cases the penalized algorithms \code{gfpop3} and \code{gfpop4} with $\beta = \beta_0=2\sigma^2\log(n)$ are better. For the corrupted scenario, we need the robust loss of algorithms \code{gfpop2} and \code{gfpop4} to get a much better MSE than other approaches. To confirm the benefit of using a penalized approach in Student case, we plot the distribution of the MSE for the five best algorithms in Figure~\ref{violinPlot}.

\begin{table}[ht!]
\begin{tabular}{|c|c|c|c|c|c|c|c|c|}
\hline
Iso-step& linear &\code{isoreg}&\code{reg}\underline{\hspace{0.2cm}}\code{1d}&\code{reg}\underline{\hspace{0.2cm}}\code{1d}&\code{gfpop1} &\code{gfpop2}&\code{gfpop3} &\code{gfpop4}\\
simulations& fit & $\ell_2$ & $\ell_1$ & $\ell_2$&$\beta = 0$ &$\beta = 0$ & $\beta = \beta_0$ & $\beta = \beta_0$\\
 & && &  & $\ell_2$ & $K = 3\sigma$& $\ell_2$ & $K = 3\sigma$ \\ \hline
Gauss&& & & & & & & \\ 
MSE& 8.27 & 0.635 & 0.632 & 1.34 & 1.21 & 0.842 & \bf 0.358 & \bf 0.470\\
(SD)&(0.022) &(0.11) &(0.10) &(0.30) &(0.78) &(0.16) &(0.12) &(0.17) \\ \hline

Student&& & & & & & & \\ 
MSE& 8.27 & 0.571 & 0.569 &  0.564 & 1.09 & 0.439  & \bf 0.301 & \bf 0.201 \\
(SD)&(0.024) &(0.15) &(0.15) &(0.14) &(1.0) &(0.13) &(0.15) &(0.073) \\ \hline

Corrupted&& & & & & & & \\ 
MSE& 304 & 300 & 300 & 30.1  & 297 & \bf 3.57 & 301 & \bf 3.17 \\
(SD)&(7.6) &(7.3) &(7.2) &(3.5) &(11) &(0.53) &(7.3) &(0.59) \\ \hline
\end{tabular}
\caption{\label{isosteptab} Mean squared errors MSE $= \frac{1}{n}\sum_{i=1}^n(\hat{s}_i - s_i)^2$ for different algorithms on step-wise simulations with its empirical standard deviation (SD). We consider three types of noise: Gaussian, Student and corrupted.}
\end{table}

We also compare the ability of the different methods to estimate the number of steps. The average number of steps over $100$ simulations is reported in Table~\ref{isosteptab2}. Only the penalized algorithms are able to recover the true number of steps ($10$). 
The choice of a good penalty in isotonic simulations is an area of ongoing research in statistics  \citep{gao2017estimation}.

\begin{table}[ht!]
\begin{tabular}{|c|c|c|c|c|c|c|c|}
\hline
Iso-step &\code{isoreg}&\code{reg}\underline{\hspace{0.2cm}}\code{1d}&\code{reg}\underline{\hspace{0.2cm}}\code{1d}&\code{gfpop1} &\code{gfpop2}&\code{gfpop3} &\code{gfpop4}\\
simulations & $\ell_2$ & $\ell_1$ & $\ell_2$&$\beta = 0$ &$\beta = 0$ & $\beta = \beta_0$ & $\beta = \beta_0$\\
 & & &  & $\ell_2$ & $K = 3\sigma$& $\ell_2$ & $K = 3\sigma$ \\ \hline

Gauss&&  & & & & & \\ 
$\hat{D}$&66.8 &66.7 &59.0 &69.1 &66.7 &10.0 &10.0\\ 
(SD)&(7.0) &(7.0) &(6.0) &(7.6) &(7.6) &(0) &(0) \\ \hline

Student&& & & & & & \\ 
$\hat{D}$& 69.6 & 69.4 &63.2 &70.5 &71.0 &10.1 &10.0\\
(SD)&(7.2) &(7.2) &(6.8) &(7.9) &(8.2) &(0.24) &(0) \\ \hline

Corrupted& & & & & & & \\ 
$\hat{D}$&40.9 &40.8& 47.8 &41.5 &61.6 &11.2 &10.0\\ 
(SD)&(5.2) &(5.2) &(5.9) &(5.6) &(7.6) &(0.95) &(0.14) \\ \hline

\end{tabular}
\caption{\label{isosteptab2} Mean number of segments over $100$ simulations with $10^4$ data-points for different algorithms on step-wise simulations. We consider three types of noise: Gaussian, Student and corrupted.}
\end{table}

\end{document}